\def\year{2021}\relax
\documentclass[runningheads]{llncs}
\usepackage{times}  % DO NOT CHANGE THIS
\usepackage{url}  % DO NOT CHANGE THIS
\usepackage{graphicx} % DO NOT CHANGE THIS
\usepackage{graphicx}  % DO NOT CHANGE THIS
\usepackage{caption} % DO NOT CHANGE THIS AND DO NOT ADD ANY OPTIONS TO 
\usepackage[T1]{fontenc}

\usepackage[table]{xcolor}

\usepackage{amsmath,amssymb,amsfonts,wrapfig,cite}
\usepackage{todonotes}
\newcommand{\lstar}{L*}
\newcommand{\rmax}{\mathbb{R}_{\mathrm{max}}}
\newcommand{\real}{\mathbb{R}}
\newcommand{\nat}{\mathbb{N}}
\newcommand{\set}[1]{\left\{#1\right\}}

\usepackage{algorithm}
\usepackage[noend]{algpseudocode}
\algrenewcommand\algorithmicindent{1em}%
\newcommand{\Break}{\State \textbf{break}}
\newcommand{\MyReturn}[1]{\State \textbf{return} #1}
\newcommand{\abs}[1]{\left|#1\right|}
\newcommand{\semiring}{\mathbb{S}}

\newcommand{\dimw}{\mathrm{dim}_w}
\newcommand{\scale}{\mathrm{scale}}
\newcommand{\norm}[1]{\|#1\|}
\newcommand{\height}{\mathrm{height}}

\newcommand{\defend}{\hfill$\square$}

\newcommand{\banme}[2]{\stackrel{#1}{\stackrel{\vee}{#2}}}
\newcommand{\lang}{\mathrm{Lang}}

\newcommand{\ext}{{\mathrm{ext}}}

\usepackage{kbordermatrix}

\renewcommand{\kbldelim}{(}% Left delimiter
\renewcommand{\kbrdelim}{)}% Right delimiter

\usepackage{multirow}
\usepackage{subcaption}
\usepackage{tikz}
\usepackage{tikz-uml} 
\usetikzlibrary{shapes,snakes}

\usepackage{multicol}

\algdef{SE}[DOWHILE]{Do}{doWhile}{\algorithmicdo}[1]{\algorithmicwhile\ #1}%

\makeatletter
\newcommand{\pushright}[1]{\ifmeasuring@#1\else\omit\hfill$\displaystyle#1$\fi\ignorespaces}
\newcommand{\pushleft}[1]{\ifmeasuring@#1\else\omit$\displaystyle#1$\hfill\fi\ignorespaces}
\makeatother

\title{
Learning Weighted Finite Automata \\over the Max-Plus Semiring and its Termination
}
\author{
Takamasa Okudono\inst{1} 
\and
Masaki Waga\inst{3}
\and
Taro Sekiyama\inst{2, 4}
\and
Ichiro Hasuo\inst{2, 4}
}
\institute{
Independent Researcher, Kyoto, Japan
\and
National Institute of Informatics, Tokyo, Japan
\and
Graduate School of Informatics, Kyoto University, Kyoto, Japan
\and
The Graduate University for Advanced Studies (SOKENDAI), Tokyo, Japan
}
\newcommand{\gennote}[3]{\todo[linecolor=#2,inline,backgroundcolor=#2!25,bordercolor=#2]{#3: #1}}
\newcommand{\inlinetodo}[1]{\todo[inline]{#1}}
\newcommand{\TO}[1]{\gennote{#1}{green}{TO}}
\newcommand{\ih}[1]{{\gennote{#1}{purple}{IH}}}
\newcommand{\mw}[1]{\gennote{#1}{orange}{MW}}

\spnewtheorem{mytheorem}{Theorem}{\bfseries}{\itshape} 
\spnewtheorem{mylemma}[mytheorem]{Lemma}{\bfseries}{\itshape}
\spnewtheorem{myproposition}[mytheorem]{Proposition}{\bfseries}{\itshape}
\spnewtheorem{mysublemma}[mytheorem]{Sublemma}{\bfseries}{\itshape}
\spnewtheorem{mycorollary}[mytheorem]{Corollary}{\bfseries}{\itshape}
\spnewtheorem{myfact}[mytheorem]{Fact}{\bfseries}{\itshape}
\spnewtheorem{mynotation}[mytheorem]{Notation}{\bfseries}{\rmfamily}
\spnewtheorem{myremark}[mytheorem]{Remark}{\bfseries}{\rmfamily}
\spnewtheorem{myexample}[mytheorem]{Example}{\bfseries}{\rmfamily}
\spnewtheorem{myassumption}[mytheorem]{Assumption}{\bfseries}{\rmfamily}
\spnewtheorem{myconvention}[mytheorem]{Convention}{\bfseries}{\rmfamily}
\spnewtheorem{mydefinition}[mytheorem]{Definition}{\bfseries}{\rmfamily}
\spnewtheorem{myrequirements}[mytheorem]{Requirements}{\bfseries}{\rmfamily}
\spnewtheorem{myproblem}[mytheorem]{Problem}{\bfseries}{\rmfamily}
\spnewtheorem*{myproof}{Proof}{\itshape}{\rmfamily}

\usepackage{cleveref}	% required for `\cref' (yatex added)
\usepackage{version}	% required for `\comment' (yatex added)

\usepackage{floatflt}

\usepackage{comment}
\specialcomment{auxproof}
{\mbox{}\newline\textbf{BEGIN: AUX-PROOF}\dotfill\newline}
{\mbox{}\newline\textbf{END: AUX-PROOF}\dotfill\newline}
\excludecomment{auxproof}

\begin{document}

\maketitle

\begin{abstract}
Active learning of finite automata has been vigorously pursued for the purposes of analysis and explanation of black-box systems. 
% The classic \lstar\ algorithm by Angluin has seen many extensions; one quantitative extension is by Balle and Mohri that learns automata weighted by the real (times-plus) semiring. 
% A notable direction of extension is to make automata weighted and their languages quantitative---such as Balle and Mohri's extension that learns automata weighted by the real semiring (with the usual addition and multiplication of real numbers). 
% Another work in this line is by de Heerdt but we don't mention it here.
In this paper, we study an \lstar-style learning algorithm for weighted automata over the \emph{max-plus semiring}. The max-plus setting exposes a ``consistency'' issue in the previously studied semiring-generic extension of \lstar: we show that it can fail to maintain consistency of tables, and can thus make equivalence queries on obviously wrong hypothesis automata. 
 We present a theoretical fix by a mathematically clean notion of \emph{column-closedness}. 
 We also present a nontrivial and reasonably broad class of weighted languages over the max-plus semiring in which our algorithm terminates.
%  Our algorithm is applicable to general semirings.
%   Our theoretical fix makes max-plus weighted automata learning much more efficient, as is demonstrated by our experimental results.
% In this paper, we introduce the first \lstar-style  learning algorithm for weighted automata over the \emph{max-plus semiring}. Its practical relevance is shown experimentally---the choice of a right semiring (such as real vs.\ max-plus) enhances the performance of learning. 

% Perhaps more interestingly, we find that the previously studied ``semiring-generic'' extension of \lstar\ learns wrong automata for the max-plus semiring. We present a theoretical fix  by formulating the notion of   \emph{column-closedness}, which enables a
%  new  semiring-generic algorithm that we prove to be correct. Our max-plus algorithm is obtained as its instance.
\end{abstract}

% \todo[size=\tiny,inline]{This uses AAAI20 style file.  Update it when the new one is available.}
% \todo[size=\tiny,inline]{7 pages technical things + 2 pages references}

\section{Introduction}\label{sec:intro}
%\noindent\myparagraph{Background} 
\paragraph{Background} 
The topic of \emph{active  automata learning} has seen years of extensive study. The initial success is brought by Angluin's \emph{\lstar\ algorithm}~\cite{DBLP:journals/iandc/Angluin87}, which repeatedly issues so-called \emph{membership queries}, organize their answers in a table, and use the table to construct a deterministic finite automaton (DFA). A black-box system can be thought of as an oracle that answers membership queries. This way, the algorithm constructs a DFA that serves as a white-box surrogate of the black-box system.

The mathematical cleanness of the \lstar\ algorithm has led to a number of extensions. A notable one is a quantitative \emph{weighted} extension~\cite{DBLP:conf/cai/BalleM15}, where membership queries are answered by real numbers instead of Boolean values (accept or reject). The algorithm is best understood in linear algebra terms, where the notion of table is alternatively described by (a finite subblock of) the so-called \emph{Hankel matrix}. The output of the algorithm is a \emph{weighted finite automaton} (WFA), where the notions of initial state, final state, and transition are all weighted by real numbers.

WFA is a general notion that is parametric in the choice of an underlying \emph{semiring}. A semiring $\mathbb{S}$ is an algebraic structure with operations $\oplus$ and $\otimes$, subject to certain equation axioms. In a WFA over a semiring $\mathbb{S}$, weights are taken from its underlying set; they are aggregated along a path (i.e.\ a sequence of transitions) by $\otimes$; and these path weights are ``summed up'' by $\oplus$  over different paths. 

The aforementioned work~\cite{DBLP:conf/cai/BalleM15} is over the \emph{real semiring}, which is carried by the set of real numbers and whose semiring operations $\oplus$ and $\otimes$ are the usual addition ($+$) and multiplication ($\times$) of real numbers. Another well-known example of a semiring is the \emph{max-plus semiring}, a member of the family called \emph{tropical semirings}, where $\oplus$ takes the maximum of real numbers and $\otimes$ is the usual addition $+$. WFAs over tropical semirings have applications in multiple domains, from computer science~\cite{DBLP:conf/soda/AminofKL09,DBLP:journals/tocl/ChatterjeeDH10} to control theory~\cite{612036,SCL:Schutter00}. 
One straightforward interpretation of the max-plus semiring is that weights are rewards: they are accumulated along a path; and a strategy chooses the path that gives the maximum reward.

%\vspace{.5em}\noindent\myparagraph{Active WFA Learning over the Max-Plus Semiring}
\paragraph{Active WFA Learning over General Semirings}
In this paper, we study active WFA learning over the max-plus semiring. \lstar-style WFA learning  over general semirings is recently studied in~\cite{DBLP:conf/fossacs/HeerdtKR020}. We found that the algorithm used there---a natural adaptation of the algorithm in~\cite{DBLP:conf/cai/BalleM15}---has a ``consistency'' issue.
We show that this issue causes the naive extension of~\cite{DBLP:conf/fossacs/HeerdtKR020,DBLP:conf/cai/BalleM15} for the max-plus semiring to stall, and clarify why the extensions ours and \cite{DBLP:conf/fossacs/HeerdtKR020} progresses the learning.
% , one that makes the tables bigger than necessary, increases the number of queries and thus potentially harms the performance.
% can give a wrong WFA $A$. Here ``wrong'' (we say \emph{unfaithful}) means that $A$'s weight for a word $w$ can differ from the weight given  by the membership oracle for $w$. 
 We present a theoretical fix---by a mathematically clean notion of \emph{column-closedness}---and an algorithm that implements the fix. 
% We examine the cause of this anomaly, and identify a subtle issue with  the max-plus semiring. We present a fix that is based on a novel yet clean algebraic notion of \emph{column-closedness}. We show that the fix applies to general semirings---beyond the max-plus one---ensuring the faithfulness of learned WFAs.

% \vspace{.5em}
% \noindent\myparagraph{Outline of \lstar-Style Algorithms}  
\paragraph{Outline of \lstar-Style Algorithms}  
We first illustrate the issue in max-plus WFA learning that we identified. It arises  from the notion of \emph{consistency} of \lstar\ tables. Towards its explanation, we first give an outline of \lstar-style algorithms. 

The problem of active WFA learning in the style of \lstar\ is formulated as follows. Here $\mathbb{S}$ is the underlying semiring. Some notions here will be formally introduced later.
\begin{myproblem}[active WFA learning]\label{problem:WFALearning}
Let $f\colon \Sigma^{*}\to \mathbb{S}$ be a function (called a \emph{weighted language}); it is hidden from us. 
We are given the following two oracles.
\begin{itemize}
 \item  
%(1) 
A \emph{membership oracle} $m$ that takes a word $w\in\Sigma^{*}$ and returns $f(w)\in\mathbb{S}$.
 \item 
%(2) 
An \emph{equivalence oracle} $e$ that takes a WFA $A'$ over $\mathbb{S}$, and
\begin{itemize}
 \item 
 returns the symbol $\mathtt{Eq.}$ if the weighted language $f_{A'}$ induced by the WFA $A'$ coincides with $f$; and
 \item 
 returns $w$ otherwise, where $w$ is a word such that $f_{A'}(w)\neq f(w)$. This $w$ is called a \emph{counterexample}.
\end{itemize}
\end{itemize}
The desired output is a WFA $A'$ that mimics $f$ (i.e.\ $f_{A'}(w) = f(w)$ for any word $w\in \Sigma^*$, using the notation introduced later). 
\end{myproblem}

\begin{auxproof}
 \begin{figure}[tbp]
 \small
 \begin{tikzpicture}[node distance=0.3cm, minimum height=1cm,scale=0.95,every node/.style={transform shape}]
  \node[draw,rectangle, align=center] (node1) {(1) Initialize\\ Hankel\\ mask};
  \node[draw,rectangle, align=center, node distance=1.2cm, below=of node1] (node2) {(2) Expand\\ Hankel\\ mask};
  \node[draw,rectangle, align=center, node distance=0.5cm, right=of node2] (node3) {(3) Fill\\ Hankel\\ subblock};
  \node[draw,diamond, align=center, right=of node3, minimum width=1cm,minimum height=1cm] (node4) {(4)};
  \node[draw,rectangle, align=center, right=of node4] (node5) {(5) Make\\ autom. $A'$};
  \node[draw,rectangle, align=center, node distance=1.6cm, above=of node3,minimum width=1cm,minimum height=1cm] (node6) {(6) Expand\\Hankel mask\\with $w$};
  \node[draw,diamond, align=center, node distance=0.5cm,right=of node5,minimum width=1cm,minimum height=1cm] (branch) {};
  \node[draw,rectangle, node distance=1.2cm, align=center, above=of branch] (end) {Return\\the autom.};

  % umlactor is not a node and it seems its usage is different from the usual shapes. We use dummy nodes to specify the position
  \def\memOracleX{2.1}
  \def\oracleY{-4.2}
  \node (memOracle) at (\memOracleX,\oracleY) {};
  \umlactor[x=\memOracleX,y=\oracleY,scale=0.8,below=0.1cm]{membership oracle $m$}
  \def\eqOracleX{6.9}
  \node (eqOracle) at (\eqOracleX,\oracleY) {};
  \umlactor[x=\eqOracleX,y=\oracleY,scale=0.8,below=0.1cm]{equivalence oracle $e$}

  \path[->,thick]
  (node1) edge (node2)
  (node2) edge (node3)
  (node3) edge (node4)
  (node4) edge[bend right=60] node[below,align=center]{$H_{(P,S)}$ is\\column- and\\ row-closed} (node5)
  (node4) edge[bend right=40] node[above right,align=center]{$H_{(P,S)}$ is \emph{NOT}\\column- or row-closed} (node2)
  (node5) edge (branch)
  (branch) edge node[right,align=center] {$e(A')$\\ $= \mathtt{Eq.}$} (end)
  (branch) edge[bend right=25] node[above,align=center] {$e(A') = w$\\ (counter-ex.)\\\quad}(node6)
  (node6) edge[bend right=40] (node3)
  ;
  \path[->,dashed,thick]
  (node3) edge[bend right=15] node[left, align=center]{What's the\\value for $w$?} (memOracle)
  (memOracle) edge[bend right=15] node[right] {$m(w)$} (node3)
  (branch) edge[bend right=15] node[left,align=center]{Do we have\\ $f = f_{A'}$?} (eqOracle)
  (eqOracle) edge[bend right=15] node[right] {$e(A')$} (branch)
  ;
 \end{tikzpicture}
 \caption{Our \lstar-style  algorithm for the max-plus semiring. Dashed lines indicate interaction with the oracles. Rudimentarily speaking, a Hankel mask is the size of a table, and a Hankel subblock $H_{(P,S)}$ is a table. The existing algorithms check only row-closedness and not column-closedness.}%
 \label{fig:lstar}
 \end{figure}
\end{auxproof}
\begin{figure}[tbp]
 \small
 \centering
\scalebox{.7}{ \begin{tikzpicture}[node distance=0.3cm, minimum height=1cm,scale=0.95,every node/.style={transform shape}]
  \node[draw,rectangle, align=center] (node1) {(1) Initialize\\ the table size};
  \node[draw,rectangle, align=center, node distance=1.2cm, below=of node1] (node2) {(2) Expand\\ the table size};
  \node[draw,rectangle, align=center, node distance=0.5cm, right=of node2] (node3) {(3) Fill\\ the table};
  \node[draw,diamond, align=center, right=of node3, minimum width=1cm,minimum height=1cm] (node4) {(4)};
  \node[draw,rectangle, align=center, right=of node4] (node5) {(5) Make\\ hypothesis\\ autom. $A'$};
  \node[draw,rectangle, align=center, node distance=1.6cm, above=of node3,minimum width=1cm,minimum height=1cm] (node6) {(6) Expand\\the table size\\with $w$};
  \node[draw,diamond, align=center, node distance=0.5cm,right=of node5,minimum width=1cm,minimum height=1cm] (branch) {};
  \node[draw,rectangle, node distance=1.2cm, align=center, above=of branch] (end) {Return\\the autom.};

  % umlactor is not a node and it seems its usage is different from the usual shapes. We use dummy nodes to specify the position
  \def\memOracleX{2.3}
  \def\oracleY{-4.3}
  \node (memOracle) at (\memOracleX,\oracleY) {};
  \umlactor[x=\memOracleX,y=\oracleY,scale=0.8,below=0.1cm]{membership oracle $m$}
  \def\eqOracleX{7.5}
  \node (eqOracle) at (\eqOracleX,\oracleY) {};
  \umlactor[x=\eqOracleX,y=\oracleY,scale=0.8,below=0.1cm]{equivalence oracle $e$}

  \path[->,thick]
  (node1) edge (node2)
  (node2) edge (node3)
  (node3) edge (node4)
  (node4) edge[bend right=60] node[below,align=center]{the table is\\closed and\\ consistent} (node5)
  (node4) edge[bend right=40] node[above right,align=center]{the table is \emph{not}\\closed or consistent} (node2)
  (node5) edge (branch)
  (branch) edge node[right,align=center] {$e(A')$\\ $= \mathtt{Eq.}$} (end)
  (branch) edge[bend right=25] node[above,align=center] {$e(A') = w$\\ (counter-ex.)\\\quad}(node6)
  (node6) edge[bend right=40] (node3)
  ;
  \path[->,dashed,thick]
  (node3) edge[bend right=15] node[left, align=center]{What's the\\weight for $w$?} (memOracle)
  (memOracle) edge[bend right=15] node[right] {$m(w)$} (node3)
  (branch) edge[bend right=15] node[left,align=center]{$f = f_{A'}$?} (eqOracle)
  (eqOracle) edge[bend right=15] node[right] {$e(A')$} (branch)
  ;
 \end{tikzpicture}
} \caption{\lstar-style  algorithms, an outline.
Dashed lines indicate interaction with the oracles. Algorithms differ in what exactly they require in ``closedness'' and ``consistency.''}%
 \label{fig:lstar}
 \vspace{-2em}
\end{figure}
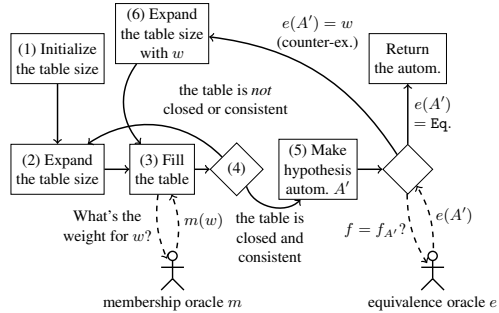

\begin{wrapfigure}[5]{r}{2cm}
\footnotesize
%\!\!\!

\vspace{-2.5em}
\begin{math}
     \begin{tabular}{l|ll}
$P\backslash S$ & $\epsilon$ & $a$ 
%& $(P\set{a,b})\backslash S$ & $\epsilon$ & $a$ & comb.
\\\hline
$\epsilon$ & 1.3   &   2.6
% & $aa$ & 34 & 42 & $8\otimes a$ 
\\
$a$& 2.6   &   3.4  
%& $aba$ & 40 & 48 & $14\otimes a$ 
\\
$ab$& 3.5   &   4.0 
% & $abb$ & 44 & 49 & $9\otimes ab$ 
\\
$b$& 2.8   &   3.0 
% & $ba$ & 30 & 39 & $13\otimes \epsilon \oplus 2\otimes b$ 
\\
% &   &    & $bb$ & 37 & 42 & $2\otimes ab$ 
    \end{tabular}
\end{math}
\end{wrapfigure}
An \lstar-style algorithm solves the problem in the following way (Fig.~\ref{fig:lstar}). It issues multiple membership queries, recording the oracle's answers in a table. An example of a table is shown on the right (for real-number weights), where rows and columns are indexed by words $p\in P$ and $s\in S$ (where $P,S\subseteq \Sigma^{*}$). The entry at $(p,s)$ is the weight $f(ps)$ returned by the membership oracle. Such tables amount mathematically to   subblocks of the so-called \emph{Hankel matrix}.

Once the table is \emph{closed} and \emph{consistent} in a suitable sense, the table can induce a WFA (Step 5 in Fig.~\ref{fig:lstar}). The resulting WFA is examined in an equivalence query. If the oracle's answer is not $\mathtt{Eq.}$, the oracle returns a counterexample word $w$ as a witness of non-equivalence, which is in turn used for further expanding the table.  
% We note that it is rare that a complete equivalence oracle is provided in real-world usecases; one would therefore rely on approximate methods such as testing~\cite{DBLP:conf/aaai/OkudonoWSH20}. 

% \vspace{.5em}
% \noindent\myparagraph{Closedness and Consistency in \lstar} 
\paragraph{Closedness and Consistency in \lstar} 
The technical core of \lstar-style algorithms is  expanding a table so that it is closed and consistent (Step 2-4 in Fig.~\ref{fig:lstar}). The bottom line here is that a row of a table corresponds to a state in a WFA.
% and the $s$-th entry of the row shows the weight obtained by reading the word $s$ from the corresponding state. 
\emph{Closedness} is about having \emph{enough rows}: it requires that, from each row, its ``successor states'' should be representable using the existing states (i.e.\ rows). Here \emph{representable} means \emph{identical} in the original \lstar\ for DFAs, and \emph{linearly dependent} in the extensions for WFAs. 
 \emph{Consistency} of a table requires that such representations are consistent with the transition behaviors.
 Another way of putting it is that consistency means that the ``successor state'' is unique for each row.
%  Another common way of putting it is that consistency means \emph{no ambiguity in transitions} between states (i.e.\ rows). 

% find their counterparts as rows in the table. That is, to be specific, all the successor states should be expressible using the existing states (i.e.\ rows)---where \emph{expressible} means \emph{identical} in the original \lstar\ and \emph{linearly dependent} in se

% as linear combinations of the existing states (i.e.\ rows). 

Different \lstar-style algorithms use different mechanisms to ensure consistency. 
The original \lstar~\cite{DBLP:journals/iandc/Angluin87} conducts explicit check of both closedness and consistency. It adds rows if the table is not closed (i.e.\ not enough rows). When the table is not consistent, it means that successor states are ambiguous in the currently exposed transition relations.  To resolve this ambiguity, the algorithm adds new columns.
% it means that the currently exposed transition behaviors---in the form of entries of a row (i.e.\ a state)---are not enough to resolve ambiguity of transitions. Therefore the algorithm adds new columns.

In the semiring-weighted setting, consistency is  hard to check directly, due to the use of linear combinations for state representation. Therefore the WFA learning algorithms in~\cite{DBLP:conf/fossacs/HeerdtKR020,DBLP:conf/cai/BalleM15} employ alternative indirect measures inspired by or similar to~\cite{DBLP:journals/iandc/MalerP95}. These measures consist of clever handling of counterexamples (Step 6 in Fig.~\ref{fig:lstar}). 

\paragraph{Consistency Issue in the Max-Plus WFA Learning}
For the max-plus semiring, we first tried an implementation of the semiring-generic algorithm in~\cite{DBLP:conf/fossacs/HeerdtKR020}, and unexpectedly found that some hypothesis automata $A'$ are \emph{unfaithful}, in the sense that they return a weight $f_{A'}(w)$ that is different from the weight $m(w)$ that the membership oracle answered previously.\footnote{We note that the focus of the paper~\cite{DBLP:conf/fossacs/HeerdtKR020}  is on the termination of learning (i.e.\ whether tables can be closed in finite steps), which is orthogonal to ours. It does not claim the faithfulness of hypothesis automata; thus this paper does not show any of their results to be wrong. } 
 Here \emph{hypothesis automata} refer to those in Step 5 in Fig.~\ref{fig:lstar}, that is, those which are induced by tables that may not yet have passed the equivalence check. 

In other words, in the algorithm from~\cite{DBLP:conf/fossacs/HeerdtKR020}, some hypothesis automata may be unfaithful to the tables that induce them. We found its cause to be in the \emph{failure of consistency}, which in turn is  because  the notion of linear (in)dependence is fragile  for the max-plus semiring. For example, even if a row vector $r_{*}$ is linearly \emph{independent} of row vectors $r_{1},\dotsc, r_{m}$, the vector $r_{m}$ can be linearly \emph{dependent} on $r_{*}, r_{1}, \dotsc, r_{m-1}$. 

This potential unfaithfulness of hypothesis automata $A'$ does not mean that the algorithm in~\cite{DBLP:conf/fossacs/HeerdtKR020} can give a wrong final answer: in fact, words $w$ such that $f_{A'}(w)\neq m(w)$ will be eventually identified by the equivalence oracle $e$ as counterexamples, and $A'$ will be fixed accordingly. 
However, complete equivalence oracles are rare in reality, which makes this ``ultimate  correctness'' argument unrealistic. 
% However, in reality, complete equivalence oracles are usually unavailable and one would rely on approximate methods such as testing~\cite{DBLP:conf/aaai/OkudonoWSH20}.  Therefore, this ``ultimate correctness'' argument is unrealistic.
Given the possibility of using a hypothesis automaton $A'$ itself as a surrogate model (without $A'$ passing the equivalence check), $A'$ being unfaithful even to previous membership queries should better be avoided. Moreover, the algorithm in~\cite{DBLP:conf/fossacs/HeerdtKR020} relies on the equivalence oracle for the resolution of unfaithfulness, but (precise or approximate) equivalence queries are usually expensive.
%potentially leading to performance overhead.

Our countermeasure is to identify the notion of \emph{column-closedness} as a substitute for consistency and to enforce column-closedness in the course of \lstar-style WFA learning (which excludes potential unfaithfulness). 
% As a countermeasure, we introduce the notion of \emph{column-closedness} as a substitute for consistency. 
We call the usual closedness notion \emph{row-closedness} for distinction.  Column-closedness is a clean algebraic condition that is dual to row-closedness. This cleanness makes the notion a pleasantly robust one, allowing a simple equational proof of  the faithfulness of hypothesis automata  for an arbitrary semiring. Column-closedness can be checked as efficiently as row-closedness.

\paragraph{Class of Input Weighted Languages in which our Algorithm Terminates}
% 我々のアルゴリズムの停止性は、semiringという構造の性質の制約の弱さによって明かではありません。(Balle のアルゴリズムは、体上のベクトル空間のために停止することが知られています。)我々はその問題に取り組むために数値実験を行なったところ、停止しなさそうな具体例を見つけることができ、その本質的な例を構成して非停止性を示すことができました。
The terminating property of our algorithm is not obvious, as semirings have a mathematically weak structure (Balle's algorithm is known to terminate due to the strong structure of fields).  We performed numerical experiments to investigate the issue, and found a concrete example that did not seem to stop, which inspired us to construct an essential example to prove the non-termination of our algorithm.

We found many examples to terminate the algorithm in the experiment, and we wondered what input examples would cause the algorithm to terminate.  As a result, we found a non-trivial class of WFAs that cause the algorithm to terminate, and proved the termination.  The class is, roughly speaking, ``WFAs whose elements are only finite integer values.''  As there are many such WFAs, the class is mathematically meaningful.  The proof is also non-trivial and interesting, fully exploiting the max-plus property.

% \paragraph{Termination P}

\begin{auxproof}
 The problem setting is as follows. 
 \begin{quotation}
    Let $f\colon \Sigma^*\to \mathbb{S}$ be an (weighted) language.  If $m$ satisfies
    \begin{align}
        m(w) = f(w),
    \end{align}
    $e$ satisfies
    \begin{align}
        e(A') =
        \begin{cases}
            \mathtt{Eq.} & ; f = f_{A'}              \\
            w                   & ; f(w) \neq f_{A'}(w),
        \end{cases}
    \end{align}
    and the algorithm terminates with the output WFA $A$, then $f=f_A$ holds.
 \end{quotation}
 \ih{NB. Angluin's algorithm learns a DFA, which is \emph{not} a $\mathbf{2}$-valued WFA.l}
\end{auxproof}

% \vspace{.5em}
% \noindent\myparagraph{Contributions} 
\paragraph{Contributions} 
We summarize our contributions.
\begin{itemize}
 \item In the setting of WFAs over the max-plus semiring, we find that the WFA learning algorithm in~\cite{DBLP:conf/fossacs/HeerdtKR020} can return hypothesis automata that are unfaithful to previous membership queries (\S{}\ref{subsec:HeerdtEtAlDoesNotWork}).
 \item We introduce the notion of \emph{column-closedness} that replaces consistency. We also present a WFA learning algorithm based on it, which works for an arbitrary semiring. We prove that the learned WFA is always faithful (\S{}\ref{sec:columnClosedness}).
 \item We compared Van Heerdt's algorithm, our algorithm, and Balle's algorithm, and gave a unifying overview of when automata learning on general semirings proceeds correctly without stalling, based on the concept of column-closedness (\S{}\ref{subsec:HeerdtEtAlDoesNotWork}).
%  Van-Heerdtのアルゴリズム、我々のアルゴリズム、Balleのアルゴリズムを比較し、どのようなときに一般セミリング上のオートマトン学習がstallせず正しく進行するかということを、column-closednessの概念からの統一的な俯瞰をあたえました(3.3) 
 \item We prove that our algorithm using column-closedness does not generally terminate with a concrete example (\S{}\ref{subsec:mininalWFAs}).
 \item We find a non-trivial class of WFAs with ample examples to terminate our algorithm, and give a proof of the termination (\S{}\ref{sec:termination}).
%  \item Focusing on  the max-plus semiring, we further study  properties and extensions of the algorithm (\S{}\ref{sec:furtherOnMaxPlus}). We show  1) that our algorithm may not terminate or yield a minimal WFA and identify the reason, 2) that ``best-effort'' minimization is still possible, and 3) that we can make it noise-tolerant by relaxing equality constraints.
%  \item \todo{Revise} We implemented the max-plus instance of the proposed algorithm. Our experimental evaluation shows the relevance of underlying semirings in WFA learning (\S{}\ref{sec:expr}). 
\end{itemize}

\begin{auxproof}
 There is a paper~\cite{DBLP:journals/corr/abs-1904-02931}.

 Our contributions are:
 \begin{itemize}
    \item We show an \lstar-style automata learning algorithm for max-plus semirings which is executable by using an efficient algorithm to solve linear equation on ons-sided max-plus semirings (Section~\ref{sec:algorithm}).
    \item We provide a mathematical proof to show that the algorithm works correctly with our new concepts the \emph{column closedness} (Section~\ref{sec:theory}).
    \item We show with an example that the naive instantiation of the generalized WFA learning algorithm in~\cite{DBLP:conf/fossacs/HeerdtKR020} causes a glitch when it is applied to max-plus semirings, and clarify the reason (Section~\ref{sec:columnclosedness}).
    \item We investigate the reason of the inherent limitation of the algorithm (Section~\ref{sec:limitation}), and provide a heuristics to avoid the infinite loop, which is inherent in our max-plus \lstar, which is based on the experimental observation (Section~\ref{sec:termination}).
 \end{itemize}
\end{auxproof}

% \vspace{.5em}
\paragraph{Notations}
We use  NumPy-like notations: for a matrix $A$, 
%$A(p,s)$ is its $(p,s)$-entry, 
$A(p,:)$ is its $p$-th row and $A(:,s)$ is its $s$-th column. 
% Let $\Sigma$ be a finite set; it is called an \emph{alphabet} and its element $\sigma\in\Sigma$ is called a \emph{character}. The set of finite words over $\Sigma$ is denoted by $\Sigma^{*}$. The \emph{empty} word (of length 0) is denoted by $\epsilon$. For a word $w\in\Sigma^{*}$, its length is $|w|$, and $w_{i}\in\Sigma$ (with $i\in [1,|w|]$) denotes the $i$-th character of $w$ (hence $w = w_{1}w_{2}\dotsc w_{|w|}$). 
For a set of words $P\subseteq \Sigma^{*}$ and a character $\sigma\in\Sigma$, we define $\sigma P := \{\sigma w\mid w\in P\}\subseteq \Sigma^{*}$, and similarly $P\sigma :=\{w\sigma\mid w\in P\}$.
For a word $w$, $w_i$ stands for the $i$-th character of $w$.

\paragraph{Related Work}
Most related works have been discussed so far.
The WFA learning algorithms in~\cite{DBLP:conf/fossacs/HeerdtKR020,DBLP:conf/cai/BalleM15} are discussed in detail and compared with our algorithm in Section~\ref{subsec:HeerdtEtAlDoesNotWork}.
Here we cover other works.
\begin{auxproof}
Weighted automata over the max-plus semiring are regarded as one of the most important classes of WFAs as WFAs over the real field, and researchers has been working on the theoretical analysis on them~\cite{DBLP:journals/ijac/Krob94,DBLP:conf/lata/Daviaud20}.
\end{auxproof}
\begin{auxproof}
 In the control theory, max-plus linear systems, which is similar to weighted automata on the max-plus semiring, are investigated and applied to control and analyze production systems, in which the max-plus semiring is useful to describe the transportation time and waiting for the arrival of the matrieals~\cite{SCL:Schutter00, 612036}.
\end{auxproof}
In addition to the class of automata mentioned in \S\ref{sec:intro}, %papers mentioned elsewhere,
\lstar-style automata learning have been extended to various type of automata, such as symbolic automata~\cite{DBLP:conf/tacas/DrewsD17,DBLP:conf/birthday/MalerM17,AD18,DBLP:journals/lmcs/FismanFZ23}, symbolic weighted automata~\cite{DBLP:conf/icgi/SuzukiHYS21}, Markov decision processes~\cite{DBLP:journals/fac/TapplerA0EL21}, timed automata~\cite{DBLP:conf/formats/HenryJM20,APT20,ACZZZ20,XAZ22,DBLP:conf/cav/Waga23}, and nominal automata~\cite{DBLP:conf/popl/MoermanS0KS17}.
\begin{auxproof}
, and researches to integrate various automata learning also exist~\cite{DBLP:conf/fossacs/HeerdtKR020,DBLP:conf/lics/UrbatS20}.
Among such works, \cite{DBLP:conf/fossacs/HeerdtKR020} studied to extend the Balle and Mohri's algorithm~\cite{DBLP:conf/cai/BalleM15} to general semirings, and concluded that principal ideal domains are amenable to \lstar-style automata learning.  Their interests were in the class of semirings whose \lstar-style procedure always terminates.
On the other hand, our interests are in what kind of properties are needed for the table of \lstar-style procedure, and how to construct a WFA faithfully from the table, and are not in guaranteeing the termination.
In that sense, our work and their work are orthogonal.
\end{auxproof}
\lstar-style automata learning algorithms are also applied to approximate \emph{recurrent neural networks} for  explainability  and acceleration~\cite{DBLP:conf/aaai/OkudonoWSH20,DBLP:conf/icgi/AyacheEG18,DBLP:conf/icml/WeissGY18}.
% Potential application is in the recent line of work where automata are extracted from recurrent neural networks. \cite{DBLP:conf/aaai/OkudonoWSH20k}

% The focus of \cite{DBLP:conf/fossacs/HeerdtKR020} is termination of an \lstar-style algorithm over general semirings. It is thus orthogonal to that of the current paper. 

% We need to show the difference between~\cite{DBLP:conf/fossacs/HeerdtKR020} 

% and~\cite{DBLP:conf/lics/UrbatS20}.

\section{Preliminaries}
\label{section:preliminaries}

\subsection{The Max-Plus Semiring}\label{subsec:maxPlusSemiring}
In this paper, we use the max-plus semiring whose base set consists of real numbers. This choice is inessential and our results apply to other base sets such as $\mathbb{Z} \cup \set{-\infty}$ and $\mathbb{Q} \cup \set{-\infty}$. The notation $\rmax$ is standard in the community; see e.g.~\cite{akian2009linear}. 
Refer to~\cite{droste2009handbook} for the general theory of semirings and semimodules.
\begin{mydefinition}[the max-plus semiring $\rmax$]
    \emph{The max-plus semiring} $\rmax$ is a semiring defined as follows.
    \begin{itemize}
        \item The base set is $\real \cup \set{-\infty}$, where $-\infty$ is  a symbol denoting the negative infinity.
        \item (Addition $\oplus_{\rmax}$) $a\oplus_{\rmax} b$ is $\max_\real (a, b)$ if $a,b\in \real$; it is $a$ if $b=-\infty$; and it is $b$ if $a=-\infty$.
\begin{auxproof}
 The addition $\oplus_{\rmax}$ is defined by
        \begin{align}
            a\oplus_{\rmax} b = 
            \begin{cases}
                \max_\real (a, b) &; a, b \in \real\\
                a &; a\in \real, b=-\infty\\
                b &; b\in \real, a=-\infty\\
                -\infty &; a=b=-\infty
            \end{cases}.
        \end{align}
\end{auxproof}        
\item (Multiplication $\otimes_{\rmax}$) $a\otimes_{\rmax} b = a +_\real b$ if $a,b\in \real$; $a\otimes_{\rmax} b = -\infty$ otherwise.
% \begin{auxproof}
%   The multiplication $\otimes_{\rmax}$ is defined by
%         \begin{align}
%             a\oplus_{\rmax} b =
%             \begin{cases}
%                 a +_\real b &; a, b\in \real\\
%                 -\infty &; \text{otherwise}
%             \end{cases}.
%         \end{align}
%     \end{auxproof}
\item (Units) $0_{\rmax}=-\infty$, and  $1_{\rmax}=0_\real$.
    \end{itemize}
The subscripts $\bullet_{\rmax}$ may be omitted when they are clear from the context.
Obviously, $\rmax$ is commutative.
\defend
\end{mydefinition}

Over the max-plus semiring, 
a particular form of a system of linear equations can be solved efficiently~\cite{thebook}. The procedure consists of computing a solution candidate---it is called a \emph{principal solution}, following the convention in the community~\cite{thebook}, but we emphasize that a principal solution \emph{may not be a solution}---and checking whether the candidate is indeed a solution or not. This procedure is known to be complete: if the principal solution turns out to be not a solution, then the linear equation has no solution.

\begin{mydefinition}[principal solution]
    \label{def:principal_solution}
Consider a system of linear equations $xA=b$, where $A\in \rmax^{n\times m}$ is a coefficient matrix, $b\in \rmax^m$ is a row vector and $x\in \rmax^n$ is an indeterminate row vector. Its \emph{principal solution} is defined by
    \begin{displaymath}\label{eq:principalSol}
    \footnotesize
\begin{array}{c}
         x(i) = 
        \begin{cases}
            \min_{j\in [1,m]} ({b(j) - A(i,j)}) &\text{if } A(i,:)\neq(-\infty,\dots,-\infty)\\
            -\infty &\text{if } A(i,:)=(-\infty,\dots,-\infty)
        \end{cases}
\end{array}    \quad\text{for $i=1,\dots,n$}.
\end{displaymath}
    
%     For a matrix $A\in \rmax^{n\times m}$, $b\in \rmax^m$ and indeterminate $x\in \rmax^n$, the \emph{principal solution} of a system of linear equations $xA=b$ is \footnote{The usual definition (e.g.\ in~\cite{thebook}) assumes that no row is $(-\infty, \dots, -\infty)$.  We do not do so; instead we add the second case in~(\ref{eq:principalSol}) as a workaround.
%  % we add the workaround to evaluate $x_i$ as $-\infty$ in the solution.
% } 
%     \begin{align}\label{eq:principalSol}
%         x_i &= 
%         \begin{cases}
%             \min_{j\in [1,m]} ({b_j - a_{ij}}) &\text{if } a_i\neq(-\infty,\dots,-\infty)\\
%             -\infty &\text{if } a_i=(-\infty,\dots,-\infty)
%         \end{cases}
%     \end{align}
%     for $i=1,\dots,n$.
    \defend
\end{mydefinition}
Although the usual definition (e.g.\ in~\cite{thebook}) assumes that no row of $A$ is $(-\infty, \dots, -\infty)$, we do not do so; instead, we add the second clause as a workaround. Notice that this second clause resolves the apparent problem in $\min$ in the first clause: if $A(i,j)=-\infty$ then $b(j) - A(i,j)$ should be $+\infty$ that does not belong to $\rmax$; but the condition $A(i,:)\neq(-\infty,\dots,-\infty)$ ensures that there is some $j$ for which $b(j) - A(i,j)$ is well-defined.

% The $\min$ expression in~(\ref{eq:principalSol}) seems problematic when $a_{ij}=-\infty$, in which case $b_{j}-a_{ij}$ should be $+\infty$ that is not an element of the semiring. This is not a real problem since it is assumed that $a_i\neq(-\infty,\dots,-\infty)$---there is some $j$ such that $a_{ij}\neq-\infty$ and this $a_{ij}$ will be used for the minimum.

% The value $b_j-a_{ij}$ is evaluated to $+\infty$ temporarily, but by taking the minimum of $\set{b_j-a_{ij}\mid j=1,\dots,m}$, the value $x_i$ is finally in $\rmax$.
% The principal solution may or may not be the solution of the system of linear equations, but if it is not, then the system is unsolvable.
% Hence, the algorithm to solve a system of linear equations on the max-plus semiring is to: (1) calculate the principal solution, and (2) check if the principal solution is a solution.
We note that multiple solutions may exist for $xA=b$. It is known that, if the principal solution is a solution and there is no row filled with $-\infty$ in $A$, then it is the maximum among the solutions in the pointwise order. It is also obvious that a system of the form $XA=B$ can be solved, where  $X, A, B$ are matrices, by solving $X(k,:)\cdot A=B(k,:)$ for each $k$~\cite{Cuninghame-Green}.

% Remark that multiple solutions may exist in general.
% If the principal solution is a solution, then it is maximum among the solutions in the pointwise partial order.
% We can also solve a system of form $XA=B$ for matrices $X, A, B$, by solving $x_iA=b_i$ repeatedly~\cite{Cuninghame-Green}.

 There are several definitions known for linear independence in the max-plus semiring (they are not mutually equivalent). We find the following one suited for our purpose.
\begin{mydefinition}[weak linear (in)dependence~\cite{Cuninghame-Green}]\label{def:weakLinINdep}
 Let $M$ be  a semimodule over a commutative semiring $\semiring$, and  $X\subseteq M$ be a subset. $X$ is \emph{weakly linearly dependent} if there exists an element $x\in X$ such that $x$ can be expressed as a linear combination $\bigoplus_{\lambda}c_{\lambda}\otimes x_{\lambda}$
% \mw{If we have space, you should explicitly write the definition because this is not the standard linear combination but its semiring generalization} 
of  elements $x_{\lambda}\in X\setminus\set{x}$.  If $X$ is not weakly linearly dependent, it is said to be \emph{weakly linearly independent}.
    \defend
\end{mydefinition}

\begin{auxproof}
 It is known that, for each finitely generated semimodule,  the number of any weakly linearly independent generators is the same. This determines the \emph{weak dimension}~\cite{akian2009linear}.
 \begin{mydefinition}[weak dimension]
    For a finitely generated semimodule $M\subseteq \rmax^n$, the unique cardinality of its weakly linearly independent generators is referred to as the \emph{weak dimension}, and written as $\dimw(M)$.
    \defend
 \end{mydefinition}
\end{auxproof}

\subsection{Weighted Automata} 
\begin{mydefinition}[WFA]
    A \emph{weighted finite automaton (WFA)} over a semiring $\semiring$ is a tuple $(\Sigma, \alpha, \beta, (A_\sigma)_{\sigma \in \Sigma})$, where
    $\Sigma$ is a finite set  (\emph{alphabet}), $\alpha$ is a row vector in $\semiring^d$  (\emph{initial vector}), $\beta$ is a column vector in $\semiring^d$  (\emph{final vector}),
    and $A_\sigma$ is a matrix of size $d\times d$ called \emph{transition matrix} of $\sigma\in\Sigma$.
The size $d$ of the vectors and matrices is called \emph{the number of states} and denoted by $|A|$.
    \defend
\end{mydefinition}
% \mw{Perhaps you want to call $d$ as the dimension of the WFA instead of the number of states}

% The usual notion of language is a set of words, and it  is 
% %As deterministic finite automata induce a language, which is
%  equivalent to a function $\Sigma^*\to\set{0, 1}$. WFAs weighted in $\semiring$ induce \emph{weighted languages} $\Sigma^{*}\to\semiring$.
% , which is a function from $\Sigma^*$ to $\semiring$ as follows:
\begin{mydefinition}[configuration, weighted language induced from WFA]
    \\For a WFA $A=(\Sigma, \alpha, \beta, (A_\sigma)_{\sigma \in \Sigma})$, \emph{the configuration} of $A$ is a function $\delta_A\colon \Sigma^*\to \semiring^d$ defined by
    \begin{displaymath}
        \delta_A(w_1\dots w_n) = \alpha A_{w_1} \dots A_{w_n}.
    \end{displaymath}
    \\A WFA $A=(\Sigma, \alpha, \beta, (A_\sigma)_{\sigma \in \Sigma})$ induces a function $f_A\colon \Sigma^*\to \semiring$ defined by \\$f_A(w_1\dots w_n) = \alpha A_{w_1} \dots A_{w_n} \beta$.
    % \begin{displaymath}
    %     f_A(w_1\dots w_n) = \alpha A_{w_1} \dots A_{w_n} \beta.
    % \end{displaymath}
   This $f_{A}$ is called the \emph{weighted language} induced by $A$.

A weighted language that is induced by some WFA is said to be  \emph{rational}.
For a WFA $A$, if there is no WFA that has the same weighted language and has a smaller number of states than $A$, then  $A$ is said to be \emph{minimal}.
    \defend
\end{mydefinition}

% Weighted languages that are induced from WFAs are called \emph{rational}.
% For a WFA $A$, if there exists no WFA that has an induced weighted language equivalent to $A$ and a smaller number of states than $A$, then the WFA $A$ is said to be \emph{minimal}.
% For a WFA $A$, if there exists no WFAs of smaller number of states and of equivalent induced weighted language to $f_A$, the WFA $A$ is called \emph{minimal}.

 We use matrices whose rows and columns are  indexed  by words.
For label sets $P, S\subseteq \Sigma^*$ and a matrix $A\in \semiring^{P\times S}$, $A(p, s)\in \semiring$ is the entry at the  $p$-th row and the $s$-th column.

The \emph{Hankel matrix} of a language $L$ is a mathematical construct that is behind  \lstar-style automata learning. Its $p$-th row, for $p\in \Sigma^*$, is the left-derivative of $L$ by $p$. 
% In the automata theory, the Brzozowski derivative of languages plays an important role.
% The following \emph{Hankel matrix} lists the counterpart of the Brzozowski derivative for weighted languages, and plays an important role in the analysis of the weighted language.
\begin{mydefinition}[Hankel matrix]
    Let $\Sigma$ be an alphabet and $\semiring$ be a semiring.
    \emph{A Hankel matrix} $H$ is a matrix $\semiring^{\Sigma^*\times\Sigma^*}$ such that $H(r\sigma, s) = H(r, \sigma s)$ for any $r,s\in\Sigma^*$ and $\sigma \in \Sigma$.
A weighted language $f\colon \Sigma^*\to \semiring$ induces a Hankel matrix, denoted by $H_f\in\semiring^{\Sigma^*\times \Sigma^*}$, by $H_f(p, s) = f(p\cdot s)$ for words $p, s\in \Sigma^*$.
    \defend
\end{mydefinition}
\begin{auxproof}
 Hankel matrices are defined for usual languages (subsets of $\Sigma^*$) in the same manner, and the entries are $0$ or $1$ indicating if the word is accepted or rejected by the language.
\end{auxproof}
Hankel matrices are of infinite size, so algorithms use the following finite restrictions. In particular, Hankel subblocks correspond to the tables in \lstar-style algorithms (Fig.~\ref{fig:lstar}). 
% Since Hankel matrices are conceptual, in a sense that it contains infinite data, we consider its finite approximation by looking at the finite part.
\begin{mydefinition}[Hankel mask, Hankel subblock]
    \label{def:hankelmask}
    For an alphabet $\Sigma$, a \emph{Hankel mask} $(P, S)$ is a pair of subsets of $\Sigma^*$.
  The \emph{Hankel subblock} of a Hankel matrix $H$ masked by $(P, S)$, denoted by $H_{(P, S)}$, is the matrix $\semiring^{P\times S}$ such that $H_{(P, S)}(p, s) = H(p, s)$ for any $p \in P, s\in S$.
    \defend
\end{mydefinition}
Remark that the indexing of a Hankel subblock and the mask of the subblock does necessarily coincide.
For example, though the mask of $H_{(P\sigma, S)}$ is $(P\sigma, S)$, where $\sigma\in \Sigma$, $H_{(P\sigma, S)}$ is usually indexed by $P\times S$, which means, $H_{(P\sigma, S)}(p, s) = H(p\sigma, s)$.

\section{WFA Learning for General Semirings}\label{sec:columnClosedness}
Here we introduce our algorithm, in which our new notion of column-closedness  replaces consistency (see \S{}\ref{sec:intro}). We present the algorithm for a general semiring $\semiring$, and prove its ``correctness'' (we say \emph{faithfulness}), i.e.\ that the learned WFA respects the weights given by the membership oracle. Later in~\S{}\ref{subsec:HeerdtEtAlDoesNotWork} we show the problem with the existing algorithm from~\cite{DBLP:conf/fossacs/HeerdtKR020} how it is resolved.

% \begin{remark}
%  In our algorithm, there are two big limitations compared to the existing \lstar-style WFA learning for the real field~\cite{DBLP:conf/cai/BalleM15}.
%  First, the minimality of the output WFA $A$ is not guraranteed.  Second, the termination of the algorithm is not guaranteed.
%  Both the limitations are due to the mathematical property of max-plus semirings, and discuss in Section~\ref{sec:limitation} why they are inevitable.
%  On the second limitation on the termination, we propose a heuristics to avoid (if possible) the infinite loop and halt the algorithm in Section~\ref{sec:termination}.
% \end{remark}

\subsection{Row-Closedness and Column-Closedness}
\label{subsec:theory}
% The keys of our max-plus \lstar\ are how to check if the data in the Hankel subblock are enough to construct a WFA (Step 4), and how to construct the WFA (Step 5).
% We start with the definitions to describe the condition for the check and construction.
\begin{mydefinition}[row-closedness, column-closedness]\label{def:rowClmClosed}
Let $H\in\semiring^{\Sigma^*\times\Sigma^*}$ be a Hankel matrix, $P\subseteq \Sigma^*$ be prefix-closed,  and $S\subseteq \Sigma^*$ be suffix-closed.
% \begin{itemize}
%  \item    
  The Hankel subblock $H_{(P, S)}$ is \emph{row-closed} if the system of linear equations $X_\sigma H_{(P, S)}=H_{(P\sigma, S)}$ has a solution for any $\sigma\in \Sigma$.
   Here $X_\sigma\in \semiring^{P\times P}$ is the indeterminate matrix.
%  \item 
 Dually, $H_{(P, S)}$ is \emph{column-closed} if the system of linear equations $H_{(P, S)}Y_\sigma=H_{(P, \sigma S)}$ has a solution for any $\sigma\in \Sigma$.
   Here $Y_\sigma\in \semiring^{S\times S}$ is the indeterminate matrix.
    \defend
% \end{itemize}
\end{mydefinition}
These $X_\sigma$ and $Y_\sigma$ are used in whole this section.

 Row-closedness means that each row of $H_{(P\sigma, S)}$ is a linear combination of the rows of $H_{(P, S)}$; this is the usual condition of \emph{closedness} used in WFA learning~\cite{DBLP:conf/cai/BalleM15,DBLP:conf/fossacs/HeerdtKR020}. Our notion of column-closedness is its dual,  meaning that each column of $H_{(P, \sigma S)}$  is a linear combination of the columns of $H_{(P, S)}$. Note that, in column-closedness, the column-index set $S$ is extended \emph{backward} to $\sigma S$.

\begin{mylemma}
    \label{lem:exchange}
    Let $H_{(P, S)}$ be row-closed and column-closed, and $X_{\sigma}$ and $Y_{\sigma}$ be solutions of the equations in Def.~\ref{def:rowClmClosed}.
Then we have $X_\sigma H_{(P, S)} = H_{(P, S)}Y_\sigma$. 
\end{mylemma}
\begin{myproof}
    $X_\sigma H_{(P, S)} = H_{(P\sigma, S)} = H_{(P, \sigma S)} =  H_{(P, S)}Y_\sigma$.
\end{myproof}
Remark that $H_{(P\sigma, S)}$ and $H_{(P, \sigma S)}$ are indexed by $P\times S$.  See the remark of Definition~\ref{def:hankelmask}.
\begin{auxproof}
 \ih{drop? (Takamasa is fine)}
 The above simple lemma is a key enabler of our algorithm. It also supports the following interpretation.
 % Another interpretation of these condition is as follows: 
 Row-closedness ensures that there is a semiring homomorphism $\semiring^{P}\to\semiring^{P}$ that carries the $s$-th column to $\sigma s$-th column for each $\sigma, s$ (this follows from the first two equalities in the proof). Dually, column-closedness ensures that there is a homomorphism  $\semiring^{S}\to\semiring^{S}$
 mapping the $p$-th row to $p\sigma$-th row for each $\sigma,p$.  
 % (Note the twisted relationship between rows and columns). 
\end{auxproof}

\begin{mylemma}[shifting]%
 \label{lem:shifting}
 Let $P\subseteq \Sigma^*$ be prefix-closed, $S\subseteq \Sigma^*$ be  suffix-closed,
 and  $s,w\in\Sigma^*$ be such that $sw\in S$. If the Hankel subblock $H_{(P, S)}$ is row-closed, then we have $(X_{s_1}\dots X_{s_n} H_{(P, S)})(:, w) = H_{(P, S)}(:, sw)$, where $s=s_1s_2\dots s_n$. 
 Dually, 
 let $w, p\in \Sigma^{*}$ be such that $wp\in P$. If  $H_{(P, S)}$ is column-closed, then  we have $(H_{(P, S)} Y_{p_1}\dots Y_{p_n})(w, :) = H_{(P, S)}(wp, :)$.
 \defend
\end{mylemma}
\begin{myproof}
    We prove the first half of Lem.~\ref{lem:shifting} by induction on the length of $s$.  If $s$ is empty, it is obvious.
 If $s$ is nonempty,
 because $S$ is suffix-closed, $s_2\dots s_n w\in S$ holds, and thus, we have $(X_{s_2}\dots X_{s_n} H_{(P,S)})(:, w) = H_{(P, S)}(:, s_2\dots s_n w)$ by the induction hypothesis.
 By definition, 
 $X_{s_1}H_{(P, S)}(:, s_2\dots s_n w) = H_{(Ps_1, S)}(:, s_2\dots s_n w)$ holds, and we have
 $X_{s_1}H_{(P, S)}(:, s_2\dots s_n w) = H_{(Ps_1, S)}(:, s_2\dots s_n w) = H_{(P, s_1 S)}(:, s_2\dots s_n w) = H_{(P, S)}(:, sw)$.
 The proof for the second half of Lem.~\ref{lem:shifting} is similar.
\end{myproof}

\begin{mytheorem}[WFA construction]%
 \label{thm:main}
 Let $H$ be a Hankel matrix, $P\subseteq \Sigma^*$ be  prefix-closed, and $S\subseteq \Sigma^*$ be suffix-closed.
Assume further that  $H_{(P, S)}$ is both row-closed and column-closed. Let  
 $A=(\Sigma, \alpha, \beta, A_\sigma)$ be the following WFA: 
 $\alpha = (0,-\infty, \dots, -\infty)$ ($0$ is  at the position $\epsilon$), $\beta = H_{(P, S)}(:, \epsilon)$, for $\sigma \in \Sigma$, $A_\sigma$ is a solution of $X_\sigma H_{(P, S)} = H_{(P\sigma, S)}$.
% \@
%  \begin{itemize}
%   \item $\alpha = (
% %\banme{\text{$\epsilon$-th}}{0}, 
%  0,-\infty, \dots, -\infty)$ ($0$ is  at the position $\varepsilon$)
%   \item $\beta = H_{(P, S)}(:, \epsilon)$
%   \item For $\sigma \in \Sigma$, $A_\sigma$ is a solution of $X_\sigma H_{(P, S)} = H_{(P\sigma, S)}$
% % \footnote{Thanks to the row-closedness, there always exists such a solution $X_\sigma$ though it may not be unique.}
%  \end{itemize}
Then  for $p\in P, s\in S$,  we have $f_A(ps) = H_{(P, S)}(p, s)$.
 % \defend\mw{Probably we should not put $\square$ here.}
\end{mytheorem}
\noindent
Note that the above construction of the WFA $A$ only requires row-closedness to go through---it uses $X_{\sigma}$ but no $Y_{\sigma}$. However, for the correctness property $f_A(ps) = H_{(P, S)}(p, s)$ (that the constructed WFA returns the value recorded in the Hankel subblock), we need column-closedness. This is seen in the proof below, and the necessity of the column-closedness is shown in~\S{}\ref{subsec:HeerdtEtAlDoesNotWork} with an example. 

% We use the following lemma in the proof of prove Theorem~\ref{thm:main}.

 %% Takamasa's original version
\begin{comment}
 \begin{mytheorem}[constructing WFA]%
 \label{thm:main}
 Let $H$ be a Hankel matrix, $P\subseteq \Sigma^*$ be a prefix-closed set, and $S\subseteq \Sigma^*$ be a suffix-closed set.
 If the Hankel subblock $H_{(P, S)}$ is row-closed and column-closed, then the WFA $A=(\Sigma, \alpha, \beta, A_\sigma)$ below induces the weighted language $f_A$ that is consistent with $H_{(P, S)}$, i.e, $f_A(ps) = H_{(P, S)}(p, s)$ for any $p\in P, s\in S$.\inlinetodo{Substitute ``consistent'' to a better terminology}\mw{I forgot which word we decided to use :-P}
    \begin{itemize}
        \item $\alpha = (\banme{\text{$\epsilon$-th}}{0}, -\infty, \dots, -\infty)$
        \item $\beta = H_{(P, S)}(:, \epsilon)$
        \item $A_\sigma (\sigma \in \Sigma)$ is a solution of $X_\sigma H_{(P, S)} = H_{(P\sigma, S)}$ (the existence of the solution $X_\sigma$ is guaranteed by the row-closedness)
    \end{itemize}
    \defend
\end{mytheorem}

We need a lemma to prove it.

\begin{mylemma}[shifting]
    \label{lem:shifting}
    Let $P\subseteq \Sigma^*$ be a prefix-closed set and $S\subseteq \Sigma^*$ be a suffix-closed set.
    For $w, s\in S$, if the Hankel subblock $H_{(P, S)}$ is row-closed and $sw\in S$, then $(X_{s_1}\dots X_{s_n} H_{(P, S)})(:, w) = H_{(P, S)}(:, sw)$.
    Similarly, for $w, p\in P$, if the Hankel subblock $H_{(P, S)}$ is column-closed and $wp\in P$, then $(H_{(P, S)} Y_{p_1}\dots Y_{p_n})(w, :) = H_{(P, S)}(wp, :)$.
    \defend
\end{mylemma}
\end{comment}
\begin{auxproof}
 Remark that the Hankel subblock is still labeled by $P\times S$ even if it is multiplied by other matrices.
\end{auxproof}

%------------------------------------------------
\begin{myproof}%[Proof of Theorem~\ref{thm:main}]
 % Let $p\in P$ of length $n$ and $s\in S$ of length $m$.
 Let $|p|=n$ and $|s| =m$. 
 We write $e_{w}$ for $(-\infty,\dots,
%\banme{\text{$\epsilon$-th}}{0}
0,\dots, -\infty)$ ($0$ is at the position $w$), for $w\in \Sigma^*$.
% Thm.~\ref{thm:main} holds thanks to the following.
\vspace{-.5em}
 \begin{displaymath}
\begin{array}{rl}
         f_A(ps) %&= e_\epsilon A_{p_1}\dots A_{p_n} A_{s_1}\dots A_{s_m} H_{(P, S)} e_\epsilon^\top \\ 
        &= e_\epsilon X_{p_1}\dots X_{p_n} X_{s_1}\dots X_{s_m} H_{(P, S)}(:,\epsilon) \\
        &\stackrel{\text{Lem.\ref{lem:shifting}}}{=} e_\epsilon X_{p_1}\dots X_{p_n} 
         H_{(P, S)}(:,s)
% \kbordermatrix{
%             & \banme{\text{$\epsilon$-th}}{}&\omit &\\
%             &H_{(P, S)}(:, s) &\omit\vrule& \bigast
%             }e_\epsilon^\top
% \\
%         &= e_\epsilon X_{p_1}\dots X_{p_n} X_{s_1}\dots X_{s_m} H_{(P, S)} e_\epsilon^\top\\
%         &\stackrel{\text{Lem.\ref{lem:shifting}}}{=} e_\epsilon X_{p_1}\dots X_{p_n} \kbordermatrix{
%             & \banme{\text{$\epsilon$-th}}{}&\omit &\\
%             &H_{(P, S)}(:, s) &\omit\vrule& \bigast
%             }e_\epsilon^\top\\
 \\
        &= e_\epsilon X_{p_1}\dots X_{p_n}H_{(P, S)}e_s^\top
	 %\label{eq:ex1}
	 \\
        &\stackrel{\text{Lem.\ref{lem:exchange}}}{=} e_\epsilon H_{(P, S)}Y_{p_1}\dots Y_{p_n}e_s^\top
	 %\label{eq:ex2}
	 \\
        &=  H_{(P, S)}(\epsilon,:)\,Y_{p_1}\dots Y_{p_n}e_s^\top
	%  \\
        % &\stackrel{\text{Lem.\ref{lem:shifting}}}{=} e_\epsilon \kbordermatrix{
        %     &\\
        %     \text{$\epsilon$-th \tiny $>$}&H_{(P, S)}(p, :)\\\cline{2-2}
        %     & \bigast 
        % }e_s^\top
 \\
        &\stackrel{\text{Lem.\ref{lem:shifting}}}{=}  H_{(P, S)}(p, :)\, e_s^\top
        =H_{(P, S)}(p, s).
\end{array} 
\end{displaymath}

\vspace{-.5em}
\end{myproof}
%------------------------------------------------
\noindent We note the simplicity of the above proof. It  uses  algebraic laws from Lem.~\ref{lem:exchange}--\ref{lem:shifting}, and does not rely on any matrix constructions such as rank factorization. This is in contrast with existing proofs such as~\cite[Thm.~3]{DBLP:conf/cai/BalleM15}.

%%%%%%%%%%%%%%%%%%%%%%%%%%%%%%%%%%%%%%%%%%%%%%%%%
\subsection{Generic WFA Learning Algorithm}\label{subsec:algorithm}
%%%%%%%%%%%%%%%%%%%%%%%%%%%%%%%%%%%%%%%%%%%%%%%%%
The theoretical development in~\S{}\ref{subsec:theory} yields an \lstar-style WFA learning algorithm in a straightforward manner: we add the check of column-closedness to the algorithm in~\cite{DBLP:conf/cai/BalleM15,DBLP:conf/fossacs/HeerdtKR020}. Fortunately, column-closedness is a linear-algebraic dual to row-closedness (Def.~\ref{def:rowClmClosed}), so checking it is no harder than  row-closedness.

Our algorithm is in Algorithm~\ref{alg:ours}. It follows the outline in Fig.~\ref{fig:lstar}, where  Step 4 (check of closedness and consistency) is replaced by check of row- and column-closedness. It expands the table  (i.e.\ the Hankel mask $(P,S)$), filling the entries by issuing membership queries, until the Hankel subblock $H_{(P,S)}$ is row- and column-closed (Line~\ref{algline:extractStart}--\ref{algline:extractEnd}). 

The \textsc{Enclose-Row} procedure for checking and enforcing row-closedness is much like in the existing algorithms~\cite{DBLP:conf/cai/BalleM15}, where the $p\sigma$-th row is checked if it is a linear combination of existing rows, and if the answer is no, it is added to the row set $P$. 

 The \textsc{Enclose-Column} procedure is totally dual to   \textsc{Enclose-Row}. If the closedness check fails, a new word $\sigma s$ is added to the column set $S$. 

 The feasibility of the \textsc{Enclose-Row} and \textsc{Enclose-Column} procedures depends on whether the equation in Line~\ref{algline:rowcond} and the corresponding equation in \textsc{Enclose-Column} can be solved, that is more precisely, whether we can detect the existence of a solution. We can do so efficiently with the max-plus semiring, as we discussed in~\S{}\ref{subsec:maxPlusSemiring}. 
Once  $H_{(P,S)}$ is seen to be row- and column-closed, a WFA is constructed by the recipe in Thm.~\ref{thm:main} (Line~\ref{algline:constructWFA} in Algorithm~\ref{alg:ours}). 
When the constructed WFA is not the answer, the equivalence oracle returns a counterexample word $w$. The way in which $w$ is handled (Step 6 in Fig.~\ref{fig:lstar}) is the same as in~\cite{DBLP:journals/iandc/MalerP95,DBLP:conf/cai/BalleM15,DBLP:conf/fossacs/HeerdtKR020}, adding all  relevant suffices of $w$ to the column set $S$ (Line~\ref{algline:ctexstart}--\ref{algline:ctexend}).

As in the previous work~\cite{DBLP:conf/cai/BalleM15,DBLP:conf/fossacs/HeerdtKR020}, our algorithm terminates on some semirings such as the real semiring and PIDs.  However, it is not guaranteed to terminate on an arbitrary semiring---for example, our algorithm may diverge in learning a WFA over the max-plus semiring.
We believe that this non-termination is due to the nature of underlying semirings, instead of  the choice of an algorithm.
Even if the linear equations in \textsc{Enclose-Row} and \textsc{Enclose-Column}  are feasible (Line~\ref{algline:rowcond} and the corresponding line in \textsc{Enclose-Column}), there is no guarantee that finite repetition of them leads to a row- and column-closed table. This is indeed the case with the max-plus semiring, as we discuss in \S{}\ref{subsec:mininalWFAs}. Another issue is that the resulting WFA is not necessarily minimal (see \S{}\ref{subsec:mininalWFAs}).

% Unlike the case for the real semiring~\cite{DBLP:conf/cai/BalleM15} and for the PIDs~\cite{DBLP:conf/fossacs/HeerdtKR020}, our algorithm is not guaranteed to terminate in the max-plus semiring\footnote{We believe that this non-termination is due to the nature of underlying semirings, instead of  the choice of an algorithm, as we discuss in \S{}\ref{subsec:mininalWFAs}.  When our algorithm is applied to the real semiring or the PIDs, we can prove the termination by using the theories in~\cite{DBLP:conf/cai/BalleM15,DBLP:conf/fossacs/HeerdtKR020}.}. Even if the linear equations in \textsc{Enclose-Row} and \textsc{Enclose-Column}  are feasible (Line~\ref{algline:rowcond} and the corresponding line in \textsc{Enclose-Column}), there is no guarantee that finite repetition of them leads to a row- and column-closed table. This is indeed the case with the max-plus semiring, as we discuss in \S{}\ref{subsec:mininalWFAs}. Another issue is that the resulting WFA is not necessarily minimal (see \S{}\ref{subsec:mininalWFAs}).
% %  After all, we believe that these issues (non-termination and non-minimality) are due to the nature of  underlying semirings, instead of  the choice of an algorithm. For the max-plus semiring, we discuss potential countermeasures in \S{}\ref{sec:furtherOnMaxPlus}. 

We conclude with the following correctness theorem.
\begin{mydefinition}[faithfulness]\label{def:faithful}
 Let $m\colon \Sigma^*\to \semiring$ be a membership oracle, and \\$w^{(1)}, \dotsc, w^{(K)}$ be words (intuitively, these are the membership queries that have been issued). We say that a WFA $A$ is \emph{faithful} to $m$ on $w^{(1)}, \dotsc, w^{(K)}$ if $f_{A}(w^{(k)})=m(w^{(k)})$ for each $k\in [1,K]$.
\end{mydefinition}

\begin{mytheorem}\label{thm:ourAlgoFaithful}
 % The WFA $A_{\mathrm{ext}}$ returned by Algorithm~\ref{alg:ours} is faithful to the oracle $m$, on the words $w^{(1)}, \dotsc, w^{(K)}$ that were queried to $m$ during the algorithm's execution. \defend
 In an execution of  Algorithm~\ref{alg:ours}, each hypothesis WFA $A_{\mathrm{ext}}$ produced in Line~\ref{algline:hypothesis}  is faithful to the oracle $m$, on the words $w^{(1)}, \dotsc, w^{(K)}$ that were queried to $m$ previously during the algorithm's execution. \defend
\end{mytheorem}
\begin{myproof}
    Note first that each $w^{(k)}$ can be written as $p\cdot s$ using elements
    $p\in P$ and $s\in S$ (Line~\ref{algline:subblockrt}). We have
    \begin{math}
     f_{A}(ps) = \textsc{Subblock}(P,S)(p,s)
    \end{math}
    by Thm.~\ref{thm:main} (this uses column-closedness; see also Line~\ref{algline:constructWFA}).  The latter is equal to $m(p\cdot s)$ by Line~\ref{algline:subblockrt}. 
   \end{myproof}

% The key of our max-plus \lstar{} algorithm\mw{This is actually not an algorithm but a procedure. Do we care it?} is how to check if the data in the Hankel subblock has enough data to construct a WFA (Step 4 of Fig.~\ref{fig:lstar}), and how to construct the WFA (Step 5 of Fig.~\ref{fig:lstar}).
% Theorem~\ref{thm:main} shows the condition of the Hankel subblock we want: it has to be row-closed and column-closed.
% Hence the idea of the algorithm is to aim at expanding rows and columns of the Hankel subblock until it becomes row-closed and column-closed.

%------------------------------------------------
% \begin{multicols}{2}
\begin{algorithm}[t]
    \caption{Our WFA learning algorithm over an arbitrary semiring $\semiring$.  ``Step $n$'' refers to those in  Fig.~\ref{fig:lstar}.           Given:  $\Sigma$ (alphabet), $m$ (membership oracle), $e$ (equivalence oracle)
 }
    \label{alg:ours}
 \scriptsize
 \vspace{-2em}
 \begin{multicols}{2}
    \begin{algorithmic}[1]
        % \Statex $\bullet$ $m\colon \Sigma^*\to \semiring$: membership oracle
        % \Statex $\bullet$ $e\colon \Sigma^*\to \set{\mathtt{Eq.}}\sqcup \Sigma^*$: equivalence query answerer
        % \Statex \Comment{$e\colon \Sigma^*\to \set{\mathtt{Eq.}}\sqcup \Sigma^*$ answers equivalence queries.}
        % \Statex \hrulefill
        \Procedure{Learn}{}
        \Comment{Main procedure}
        \State $P\leftarrow \set{\epsilon}, S\leftarrow \set{\epsilon}$
        \Comment{Step 1}
        \State $A_{\mathrm{ext}} \leftarrow \textsc{Extract}(P, S)$
        \Comment{Steps 2--5} 
        \label{algline:hypothesis}
        \Loop{} \label{algline:mainloop}
        \State $\mathtt{res}\leftarrow e(A_{\mathrm{ext}})$ \label{algline:callEQQ}
        \If{$\mathtt{res} = \mathtt{Eq.}$}
        \Break
        \ElsIf{$\mathtt{res}=w$ for a word $w\in\Sigma^*$} \label{algline:ce}
        \Comment{Step 6}
        % \Statex \Comment{Found a counterexample}
        \State $p\leftarrow \text{(the longest prefix of $w$ within $P$)}$ \label{algline:ctexstart}
        \State $s\leftarrow \text{(the tail of $w$ of length $\abs{w}-1-\abs{p}$)}$ \label{algline:sets}
        \Statex \Comment{If the length is negative, it is empty}
        \State $S\leftarrow S \cup (\text{all the suffixes of $s$})$ \label{algling:add_s_by_eqq}
        \label{algline:ctexend}
        \State $A_{\mathrm{ext}} \leftarrow \textsc{Extract}(P, S)$ \label{algline:extractWFA}
        \Comment{Steps 2--5}
        \EndIf
        \EndLoop
        \MyReturn $A_{\mathrm{ext}}$
        \EndProcedure{}

        \Statex \hrulefill

        \Procedure{Extract}{$P$, $S$}
        \Do{} \label{algline:extractStart}
        \State $(P, \mathtt{updatedR?}) \leftarrow \textsc{Enclose-Row}(P, S)$
        \State $(S, \mathtt{updatedC?}) \leftarrow \textsc{Enclose-Column}(P, S)$
               \label{algline:clmclsdcheck}
        \doWhile{$\mathtt{updatedR?}$ or $\mathtt{updatedC?}$ is $\mathtt{updated}$}
         \label{algline:extractEnd}
        \MyReturn WFA $(\Sigma, \alpha, \beta, (A_\sigma)_{\sigma \in \Sigma})$ of $\textsc{H}{(P, S)}$  \label{algline:constructWFA}
        \Statex \Comment The construction of Thm.~\ref{thm:main}
        \EndProcedure{}

        \Statex \hrulefill
    % \end{algorithmic}
    % \begin{algorithmic}[1]

        \Procedure{Enclose-Row}{$P$, $S$}
        \State{$\mathtt{updated?} \leftarrow \mathtt{not\_updated}$}
        \Loop{}
        \State{$\mathtt{adding}\leftarrow \mathtt{nil}$}
        \For{$(p, \sigma)\in P\times \Sigma$}
        % \If{$\Bigl(\parbox[c]{.3\textwidth}{$\not\hspace{-.7em}\exists x\in \semiring^P$ s.t. $x\cdot\textsc{Subblock}(P, S) = \textsc{Subblock}(\set{p\sigma}, S)$}\Bigr)$}
        \If{$\not\hspace{-.2em}\exists x$ s.t. $x\cdot \textsc{H}(P, S)=\textsc{H}(\set{p\sigma}, S)$}
        \label{algline:rowcond}
        \State $\mathtt{adding} \leftarrow p\sigma$ \label{algline:set_adding}
        \Break
        \EndIf
        \EndFor
        \If{$\mathtt{adding} = \mathtt{nil}$} \label{algline:if_adding_nil}
        \Break
        \Else
        \State $P \leftarrow P \cup \set{\mathtt{adding}}$ \label{algline:add_p}
        % ;\quad \mathtt{updated?} \leftarrow \mathtt{updated}$
        \State $\mathtt{updated?} \leftarrow \mathtt{updated}$ \label{algline:set_updated}
        \EndIf
        \EndLoop{}
        \MyReturn $(P, \mathtt{updated?})$ 
        \Comment Now $H_{(P, S)}$ is row-closed
        \EndProcedure{}

        \Statex \hrulefill

        \Procedure{Enclose-Column}{$P$, $S$} \label{algline:enclose_column}
        \Statex \Comment{Omitted as it is almost the same as \textsc{Enclose-Row}}
        % \Statex \quad (the procedure dual to $\textsc{Enclose-Row}$)
        % \State{$\mathtt{updated?} \leftarrow \mathtt{not\_updated}$}
        % \Loop{}
        % \State{$\mathtt{adding}\leftarrow \mathtt{nil}$}
        % \For{$(\sigma,s)\in \Sigma\times S$}
        % \If{$\Bigl(\parbox[c]{.3\textwidth}{$\not\hspace{-.7em}\exists y\in \semiring^S$ s.t. $\textsc{Subblock}(P, S)\cdot y = \textsc{Subblock}(P, \set{\sigma s})$}\Bigr)$}
        % \label{algline:columncond}
        % \State $\mathtt{adding} \leftarrow \sigma s$
        % \Break
        % \EndIf
        % \EndFor
        % \If{$\mathtt{adding} = \mathtt{nil}$}
        % \Break
        % \Else
        % \State $S \leftarrow S \cup \set{\mathtt{adding}};\quad \mathtt{updated?} \leftarrow \mathtt{updated}$  \label{algline:SAdded}
        % % \State{$\mathtt{updated?} \leftarrow \mathtt{updated}$}
        % \EndIf
        % \EndLoop{}
        \MyReturn $(S, \mathtt{updated?})$  \Comment Now $H_{(P, S)}$ is column-closed
        \EndProcedure{}

        \Statex \hrulefill

        \Procedure{H}{$P$, $S$}
        \Comment{Returns $H_{(P, S)}$}
        % \Statex \hspace{1em}
        \MyReturn{matrix $\lambda (p, s)\in P\times S.\hspace{1em}m(p\cdot s)$}\label{algline:subblockrt}
        \Statex \Comment{Memoize answers of $m$ and implicitly process Step 3}
        \EndProcedure{}

    \end{algorithmic}
\end{multicols}
\vspace{-1.5em}
\end{algorithm}
% \end{multicols}
%------------------------------------------------

\begin{auxproof}
 %% TODO: 行番号をおきかえる
 The algorithm is shown in Algorithm~\ref{alg:WFA}.
 It starts with setting up the trivial Hankel mask $(P, S)$ (Line 2), and constructs the obvious WFA (Line 3).
 In the loop at Line 4-12, it checks if the constructed WFA $A_{\mathrm{ext}}$ is the correct answer (Line 5).
 If it is, then quit the loop and returns the constructed WFA $A_{\mathrm{ext}}$ (Line 13).
 If it is not, then expand the suffix-closed set $S$ of the Hankel mask so that the counterexample $w$ returned by the equivalence query appears in the Hankel subblock.  By the procedure at Line 9-11, the value $m(p\sigma s)$ will appear in  $p\sigma$-th row of a Hankel subblock $H_{(P\sigma, S)}$, and this row will be taken into account in the procedure $\textsc{Enclose-Row}$.

 The subprocedure $\textsc{Extract}$ (Line 14) makes the Hankel subblock determined by the given Hankel subblock $(P, S)$ row-closed and column-closed, and constructs the WFA by the Hankel subblock with the procedure given in Theorem~\ref{thm:main}.
 To make the Hankel subblock row-closed and column-closed, we applies the subprocedures $\textsc{Enclose-Row}$ and $\textsc{Enclose-Column}$ until the Hankel mask $(P, S)$ gets stable, i.e., the subprocedures do nothing to the Hankel mask (Line 15-18).
 The discussion on the termination of the loop (Line 15-18) is discussed in Section~\ref{sec:termination}.

 The subprocedure $\textsc{Enclose-Row}$ (Line 20) makes the Hankel subblock determined by the given Hankel subblock $(P, S)$ row-closed.  Since the row-closedness is equivalent to the existence of $x\in \rmax^P$ such that $xH_{(P, S)} = H_{(P, S)}(p\sigma, :)$ for any $\sigma\in \Sigma$ and $p\in P$, for all $(p, \sigma)\in P\times \Sigma$ (Line 24), the subprocedure checks the negation of the existence (Line 25), and if the negation holds, the cause of it $p\sigma$ is added to $P$ (Line 31).  This subprocedure returns the result $P$ and a flag $\mathtt{updated?}$ to let the subprocedure $\textsc{Extract}$ know if $P$ is modified or not.
 The subprocedure $\textsc{Enclose-Column}$ (Line 34) is the dual of $\textsc{Enclose-Row}$, and modifies $S$ so that the Hankel subblocks be column-closed.
 The termination of $\textsc{Enclose-Row}$ will be discussed in Section~\ref{sec:termination}.
\end{auxproof}

\subsection{Comparison with Other WFA Learning Algorithms}\label{subsec:HeerdtEtAlDoesNotWork}
% \todo[inline]{to-do: rewrite this section. The algorithm in~\cite{DBLP:conf/fossacs/HeerdtKR020} is different from Algorithm~\ref{alg:ours} in how it handles counterexamples. Say that it is faithful to the $\epsilon$-th row of the Hankel submatrix, and thus will be eventually correct. }
Van Heerdt et al.'s algorithm~\cite{DBLP:conf/fossacs/HeerdtKR020} and our algorithm have the same goal, but they differ in two techniques:
% \begin{itemize}
    % \item 
    (1) When a counterexample $w$ at Line~\ref{algline:ce} is reflected in the Hankel submatrix, our algorithm uses the \emph{Balle-inspired refinement}~\cite{DBLP:conf/cai/BalleM15}: take a suffix $s$ of the counterexample $w$, and add all the suffixes of $s$ into the column set $S$ to keep $S$ compact (Line~\ref{algline:ctexstart}-\ref{algline:ctexend}).  On the other hand, van Heerdt et al.'s algorithm uses the \emph{Maler-Pnueli-inspired refinement}~\cite{DBLP:journals/iandc/MalerP95}: add all the suffixes of the counterexample $w$ into the column set instead of Line~\ref{algline:ctexstart}-\ref{algline:ctexend} in Algorithm~\ref{alg:ours}.
    % \item
    (2) Our algorithm extends the Hankel submatrix so that it becomes column-closed (Line~\ref{algline:clmclsdcheck}), but van Heerdt et al.'s algorithm does not need this step.
% \end{itemize}

As van Heerdt et al.'s algorithm does not ensure the column-closedness at the moment of WFA construction (Line~\ref{algline:extractWFA}), the constructed WFA $A_{\mathrm{ext}}$ might be unfaithful to the previous answers of membership queries, i.e., there can be a word $w$ such that $m(w) \neq f_{A_{\mathrm{ext}}}(w)$ and the membership query of $w$ is already issued.
Even if the Hankel submatrix is not column-closed, the following property holds for the constructed WFA:
\begin{mytheorem}[constructed WFA in~\cite{DBLP:conf/fossacs/HeerdtKR020}]%
\label{thm:vanheddrt}
Let $H, P, S, A$ be the same as in Theorem~\ref{thm:main}.
If  $H_{(P, S)}$ is row-closed, then we have $f_A(s) = H_{(P, S)}(\epsilon, s)$ for $s\in S$.
% \defend\mw{Probably we should not put $\square$ here.}
\end{mytheorem}
The proof is almost the same as the proof of Theorem~\ref{thm:main}.
By this property and the way to update the column set $S$, the WFA $A_{\mathrm{ext}}$ is updated to satisfy $f_{A_{\mathrm{ext}}}(w) = f_A(w)$ for the counterexample $w$, and the learning successfully progresses.

\begin{floatingfigure}[r]{0.48\textwidth}
    \footnotesize
    % \!\!\!\!\!\!
    % \hspace{-3em}
    \vspace{-1.5em}
 \begin{align*}\label{eq:counterExTable}
        \footnotesize
        \scalebox{.8}{     \begin{tabular}{l|ll}
        $H_{(P,S)}$ & $\epsilon$ & $a$ %& $(P\set{a,b})\backslash S$ & $\epsilon$ & $a$ & comb.
        \\\hline
        $\epsilon$ & 13   &   26
        % & $aa$ & 34 & 42 & $8\otimes a$ 
        \\
        $a$& 26   &   34  
        %& $aba$ & 40 & 48 & $14\otimes a$ 
        \\
        $ab$& 35   &   40  
        %& $abb$ & 44 & 49 & $9\otimes ab$ 
        \\
        $b$& 28   &   30  
        %& $ba$ & 30 & 39 & $13\otimes \epsilon \oplus 2\otimes b$ 
        % \\
        % &   &    
        %& $bb$ & 37 & 42 & $2\otimes ab$ 
            \end{tabular}
        }   \quad \footnotesize
        \scalebox{.8}{     \begin{tabular}{l|ll|l}
        %$P\set{a,b}\quad\backslash S$
        $H_{(P\set{a,b},S)}$
        & $\epsilon$ & $a$ & comb.\\\hline
        $aa$ & 34 & 42 & $8\otimes a$ \\
        $aba$ & 40 & 48 & $14\otimes a$ \\
        $abb$ & 44 & 49 & $9\otimes ab$ \\
        $ba$ & 30 & 39 & $13\otimes \epsilon \oplus 2\otimes b$ \\
        $bb$ & 37 & 42 & $2\otimes ab$ 
            \end{tabular}
        }\end{align*}        
        \vspace{-1.5em}
    \end{floatingfigure}

One might expect that using Balle-inspired refinement and omitting to ensure the column-closedness will keep $S$ compact and accelerate the learning.
However, this ``hybrid'' method does not work actually.  
There exists a target WFA $A$ that is learnable both by our algorithm and by van Heerdt et al.'s algorithm, but makes this hybrid method diverge.  A WFA $A=(\set{a, b}, \alpha, \beta, (A_\sigma)_{\sigma\in \Sigma})$ with $\alpha_A=(6, 11, 1)$, $\beta_A=(7, 0, 6)^\top$, 
$A_a = ((2, 3, 1), (2, 0, 9), (3, 0, 8))^\top$, $A_b = ((9, 6, 2), (10, 3, 2), (8, 5, 4))^\top$
% \begin{align*}
%     A_a = \begin{pmatrix}
%         2 & 3 & 1\\
%         2 & 0 & 9\\
%         3 & 0 & 8
%     \end{pmatrix},
%     A_b = \begin{pmatrix}
%         9 & 6 & 2\\
%         10 & 3 & 2\\
%         8 & 5 & 4
%     \end{pmatrix}.
% \end{align*}
over the max-plus semiring is such an example.
Assume that the Hankel submatrix is as shown above and $A_{\mathrm{ext}}$ is the WFA constructed from the Hankel submatrix at Line~\ref{algline:mainloop} in Algorithm~\ref{alg:ours}.
% \begin{align}\label{eq:counterExTable}
% \footnotesize
% \scalebox{.8}{     \begin{tabular}{l|ll}
% $H_{(P,S)}$ & $\epsilon$ & $a$ %& $(P\set{a,b})\backslash S$ & $\epsilon$ & $a$ & comb.
% \\\hline
% $\epsilon$ & 13   &   26
% % & $aa$ & 34 & 42 & $8\otimes a$ 
% \\
% $a$& 26   &   34  
% %& $aba$ & 40 & 48 & $14\otimes a$ 
% \\
% $ab$& 35   &   40  
% %& $abb$ & 44 & 49 & $9\otimes ab$ 
% \\
% $b$& 28   &   30  
% %& $ba$ & 30 & 39 & $13\otimes \epsilon \oplus 2\otimes b$ 
% % \\
% % &   &    
% %& $bb$ & 37 & 42 & $2\otimes ab$ 
%     \end{tabular}
% }   \quad \footnotesize
% \scalebox{.8}{     \begin{tabular}{l|ll|l}
% %$P\set{a,b}\quad\backslash S$
% $H_{(P\set{a,b},S)}$
% & $\epsilon$ & $a$ & comb.\\\hline
% $aa$ & 34 & 42 & $8\otimes a$ \\
% $aba$ & 40 & 48 & $14\otimes a$ \\
% $abb$ & 44 & 49 & $9\otimes ab$ \\
% $ba$ & 30 & 39 & $13\otimes \epsilon \oplus 2\otimes b$ \\
% $bb$ & 37 & 42 & $2\otimes ab$ 
%     \end{tabular}
% }\end{align}
This Hankel matrix is row-closed, but not column-closed (the ``comb.'' column shows that it is row-closed).
In this situation, $f_{A}(ab)=H_{(P,S)}(ab,\epsilon)=35$ but $f_{A_{\mathrm{ext}}}(ab)=36$.
If the answer of the equivalence query $\mathtt{res}$ at Line~\ref{algline:callEQQ} is $ab$, then 
$s$ is set to $\epsilon$ at Line~\ref{algline:sets} and $S$ is not updated at Line~\ref{algline:ctexend}.
In total, the Hankel matrix is not updated and $A_{\mathrm{ext}}$ is also not updated.
Then, the equivalence query $e$ return $ab$ at Line~\ref{algline:callEQQ}, and this leads to the infinite loop.
The problem is that the counterexample $ab$ is not reflected in the Hankel matrix, so we have to choose either our algorithm or van Heerdt et al.'s algorithm to progress the learning.

When we turn to the real-semiring,
 the algorithm  in~\cite{DBLP:conf/cai/BalleM15} does not explicitly check column-closedness nor consistency. The resulting WFA is nevertheless faithful (Def.~\ref{def:faithful}). This is because column-closed is automatically maintained in the following way. 
% The reason why the Balle and Mohri's WFA learning algorithm for $\real$ does not need the column-closedness is that the column-closedness is automatically and implicitly satisfied.
Since $\textsc{Enclose-Row}$ adds a row $H(P\sigma, S)$ only when it is not a linear combination of the rows of $H_{(P, S)}$, the Hankel subblock $H_{(P, S)}$ is maintained to be full-rank and landscape (meaning $\abs{P}\le \abs{S}$).  Hence the subspace generated by the columns of $H_{(P, S)}$ coincides with the whole space $\real^P$, and thus in particular, any column of $H_{(P, \sigma S)}$ is expressed as a linear combination of the columns of $H_{(P, S)}$. The latter means that it is column-closed.

In the max-plus setting, however, the $\textsc{Enclose-Row}$ procedure  does not maintain row-closedness.
For illustration, observe that  the $ab$-th row of $H_{(P,S)}$ in~(\ref{eq:counterExTable}) cannot be expressed as a linear combination of the $\epsilon$-th and $a$-th rows, but
the $a$-th row is $8\otimes \text{($\epsilon$-th row)} \oplus (-9)\otimes \text{($ab$-th row)}$.  
This example shows that even if a new row cannot be expressed as a linear combination of the existing rows, adding the row can destroy the weak linear independency of the matrix (Def.~\ref{def:weakLinINdep}).

%\section{Towards Max-Plus WFA Learning}

\begin{auxproof}
 \subsection{Angluin's \lstar\ Algorithm (DFAs)}
 \ih{They use \emph{consistency}}
 % AngluinのL*アルゴリズムは、上記のオートマタラーニングの枠組みを使ってブラックボックスの正規言語Lから、Lang(A)=LとなるようなDFA Lを学習するアルゴリズムです。
 % このアルゴリズムは、membership queryを繰り返し使うことで言語Lから対応するHankel mask (P, S)とHankel subblockを構成します。どのように構成するかは\cite{Angluin}をご覧ください。

 % Hankel subblockのp行s列目はLの文字列psに対する受理非受理を示しているわけですから、
 % Hankel subblockのp行目というのは言語Lのpでの左微分$p\backslash L$の有限の部分情報であるということになります。
 % Myhill-Nerodeの定理によれば、言語の各文字列での微分たち${w\backslash L | w\in \Sigma^*}$と、その言語に対応する最小DFAの状態というのは同一視できるわけですから、
 % 構成するオートマトンA=(Q, δ, i, F)の状態QとしてHankel subblockの各行の内容$\set{r_p \in Map(S, {0, 1}) | p \in P}$を用い、状態r_p∈Qからσをよんだときの行き先$\delta(r_p, \sigma)$はr_{p\sigma}とすればHankel subblockに対応するオートマトンが作れる、というのがアルゴリズムのアイデアです。(pσはPからはみでるかもしれませんが、このようなときr_{p\sigma}は必要に応じて計算することにします)。

 % ただし、Hankel subblockからの遷移の計算のときには次の2つの問題が起こりえます。まず、p∈Pとσ∈Σについて$r_{p\sigma}$が、Hankel subblockのなかにあらわれず、$r_p\in Q$から$\sigma$をよんだときの行き先$\delta(r_p, \sigma)$が定義できないということがおこりえます。
 % これは、Hankel subblockがもしも「任意のp∈Pとσ∈Σについて、あるp'∈Pがあって、r_p=r_{p'}となる」に性質を課せば防ぐことができ、この性質をclosednessといいます。
 % 第2の問題として、p,p'∈Pでr_p=r_{p'}なるものについて、r_{p\sigma}≠r_{p'σ}となり、状態r_p(=r_{p'})からσをよんだ行き先$\delta(r_p, \sigma)$が相異なる2状態$r_{p\sigma}, r_{p'σ}$となり遷移関数δが正しく定義できないということがおこりえます。
 % これは、Hankel subblockがもしも「任意のp,p'∈Pとσ∈Σについて、r_p=r_{p'}ならばr_{p\sigma}=r_{p'\sigma}」という性質を課せば防ぐことができ、この性質をconsistencyといいます。
 % Angluinのアルゴリズムは、この2つの性質を満たすようにHankel maskとHankel subblockを拡張していくことで進行します。

 % 本稿では、semiring上のWFAでこのconsistencyについて考えることがミソになります。e
 \TO{Written something in the comment.}

 For a black box regular language $L\subseteq \Sigma^*$, the \lstar\ algorithm constructs a DFA $A$ such that $\lang(A)=L$ with limited accesses to $L$ called \emph{membership queries} and \emph{equivalence queries}~\cite{DBLP:journals/iandc/Angluin87}.
 Formally, the \lstar\ algorithm takes as input an oracle answering the membership-query $m\colon \Sigma^*\to \set{0, 1}$ and an oracle answering the equivalence-query $e: \set{\text{DFAs}}\to \set{\mathtt{Eq.}}\sqcup \Sigma^*$ and outputs a minimum DFA $A$.  The relationship between the two input oracles and the output minimum DFA $A$ is as follows:
 \begin{quotation}
    Let $L\subseteq \Sigma^*$ be a regular language.  If $m$ satisfies
    \begin{align}
        m(w) = \chi_L(w),
        \label{eq:lstarm}
    \end{align}
    and $e$ satisfies
    \begin{align}
        e(A') =
        \begin{cases}
            \mathtt{Eq.} & ; \lang(A') = L              \\
            w                   & ; w \in \lang(A')\triangle L
        \end{cases}
        \label{eq:lstare}
    \end{align}
    then the algorithm terminates, and $L=\lang(A)$ holds\footnote{For sets $A$ and $B$, $A\triangle B$ represents the symmetric difference of $A$ and $B$, which is $(A\setminus B)\cup (B\setminus A)$}.
 \end{quotation}
 Intuitively, referring to the language $L \subseteq \Sigma^*$ as the \emph{model}, the above property means that if $m$ determines the output (accepting or rejecting) of the model $\chi_L(w)$ for a word $w\in \Sigma^*$ (Equation~\ref{eq:lstarm}),
 and $e$ determines whether the language of a given DFA is equivalent to the model $L$ (Equation~\ref{eq:lstare}),
 then the algorithm correctly finds a DFA whose language is $L$.

 % \begin{figure*}[ht]
 %     \centering
 %     \includegraphics[width=.75\textwidth]{img/lstar.pdf}
 %     \caption{The flow of general \lstar\ algorithms.  Dotted lines shows the interaction with the oracles.  C.E. stands for counterexamples.}%
 %     \label{fig:lstar}
 % % to-do
 % % (1)Fontをタイムズにする(2)ワンコラムに納める,フォントがでかすぎる(3)線が細すぎる(4)人間の背が高すぎる(5)membership oracle, equivalence oracleにする(6)パワポにする？(7)Data is enoughは未定義でもいいので正確に書く(8)登場人物に名前を付ける。is this autom. correctはmとeを使って書く
 % \end{figure*}

 % L*アルゴリズムの大雑把な仕組みは図中に示してあります。まず自明なハンケルマスク$(P,S)$を用意し(1)、それを徐々に拡大させながら(2),Hankel maskの部分情報であるH(P,S)をmembership queryを発行することにより取得します(図中1)。
 The general flow of \lstar\ algorithm is shown in Figure~\ref{fig:lstar}.  It starts with making a trivial Hankel mask $(P, S)=(\set{\epsilon}, \set{\epsilon})$ (Step 1), expands the Hankel mask (Step 2), and fill the Hankel subblock by making 
 membership queries (Step 3).\inlinetodo{Define $\epsilon$ somewhere in \cref{section:preliminaries}.}
 % そのHankel submatrixが矛盾なくDFAを構成できるだけの情報を含んでいるなら(3)、そこからH(P, S)に整合するDFAを構成しequivalence queryを発行します(4)。
 If the Hankel subblock contains sufficient data to construct a DFA which is faithful to the contents of the Hankel subblock (Step 4), it constructs the DFA and make an equivalence query (Step 5).
 % もしもEquivalenceがそのqueryの答えならば、アルゴリズムは終了し(5)、そうでないならば得られたcounterexample wの情報が反映されるようにHankel maskにより拡張し(6)、ふたたびHankel maskを拡大させるステップに戻ります。
 If the atom $\mathtt{Eq.}$ is the answer of the query, it terminates, and otherwise, it expands the Hankel mask $(P, S)$ to reflect the counterexample $w$ returned by $e$ (Step 6), and goes back to Step 3 to fill the Hankel subblock.
 %% この、2種類のqueryを使いHankel maskを拡張しながらactiveに学習する手法は\lstar styleとよばれ、さまざまな形式言語にむけたvariantが作られました。

\subsection{WFA Learning for the Real Semiring}
\ih{Add the following point: They use the tweak by Maler \& Pnueli, on the use of counterexample words, so that consistency is automatically ensured
(Emphasize consistency)}

Balle and Mohri's algorithm constructs a WFA by learning a black-box rational weighted language over the real field $\real$.
Formally, Balle and Mohri's algorithm takes as input an oracle answering the membership-query $m\colon \Sigma^*\to \real$ and an oracle answering the equivalence-query $e: \set{\text{WFAs}}\to \set{\mathtt{Eq.}}\sqcup \Sigma^*$ and outputs a minimal WFA $A$.  The relationship between the two input oracles and the output minimal DFA $A$ is as follows:
\begin{quotation}
    Let $f\colon \Sigma^*\to \real$ be a rational weighted language.  If $m$ satisfies
    \begin{align}
        m(w) = f(w),
    \end{align}
    and $e$ satisfies
    \begin{align}
        e(A') =
        \begin{cases}
            \mathtt{Eq.} & ; f = f_{A'}              \\
            w                   & ; f(w) \neq f_{A'}(w)
        \end{cases}
    \end{align}
    then the algorithm terminates, and $f=f_A$ holds.
\end{quotation}
Though the general flow is the same as \lstar\ algorithm, they showed that using linear algebra makes the steps above fit to WFA learning.
The detail is deferred to Section~\ref{sec:columnclosedness} to compare with our method.

% Balle and Mohriのアルゴリズムは\lstarのフレームワークを使って、blackboxな実数体上のrational weighted languageからWFAを学習するアルゴリズムです。
% 全体的な流れ自体は変わりませんが、Balle and Mohriは線形代数を使えば、Hankel submatrixからWFAが構成できるかどうかの判定(3)や対応するWFAの構成(4)をこのタスクに適合させることができることを示しました。

\subsection{Towards the Max-Plus Semiring}\label{subsec:towardsMaxPlus}
\ih{Concrete examples can go to the appendix}

(Max-plus semiring is important. Let's do it.)

An existing work~\cite{DBLP:conf/fossacs/HeerdtKR020} introduces a general algorithm, so we may naturally want to try it. However, we find that the algorithm is not satisfactory, in the following sense of faithfulness.

\begin{mydefinition}
 A WFA is \emph{faithful} with a Hankel subblock if, ... (the entry of the Hankel subblock coincides with the WFA output).
\end{mydefinition}
Note that the entries of a Hankel subblock are the results of membership queries. Therefore, unfaithfulness means the resulting WFA exhibits behaviors that are different from the teacher.

\subsubsection{Takamasa's Original Writing}
The Balle and Mohri's WFA learning algorithm for the real field $\real$~\cite{DBLP:conf/cai/BalleM15}, which is almost the WFA learning algorithm for general ``good'' semirings~\cite{DBLP:conf/fossacs/HeerdtKR020}, is obtained just by removing $\textsc{Enclose-Column}$.
In this section, we are explaining why we have to ensure the column-closedness for the max-plus task, and why it is not needed for the real field task.

% [src/learning.rs:1047] &ls = hoge
% pres: {['a', 'b'] <> 2, [] <> 0, ['b'] <> 3, ['a'] <> 1}
% sufs: {[] <> 0, ['a'] <> 1}
% ---table below---
% 13      26
% 26      34
% 35      40
% 28      30
% ---table above---

% [src/learning.rs:1049] "ran enclose" = "ran enclose"
% [src/learning.rs:559] "started consistensify" = "started consistensify"
% [src/learning.rs:1051] "ran consistency" = "ran consistency"
% [src/learning.rs:1053] "ran add_column_when_unsolvable" = "ran add_column_when_unsolvable"
% [src/learning.rs:1055] "ran enclose_column" = "ran enclose_column"
% [src/learning.rs:1057] (&res_enclose, &res_consistensify, &res_add_column_when_unsolvable,
%  &res_enclose_column) = (
%     false,
%     false,
%     false,
%     false,
% )
% [src/learning.rs:2381] &extracted = 
% ini:    0       -13     -22     -15
% fin:    13      26      35      28
% ---trans for a---
% 8       0       -9      -2
% 16      8       -1      6
% 22      14      5       12
% 13      4       -5      2
% ---trans for a---
% ---trans for b---
% 4       -4      -10     0
% 14      6       0       7
% 23      15      9       16
% 16      8       2       9
% ---trans for b---

% thread 'learning::tests::test_extract_wfa3_problem_replay1' panicked at 'assertion failed: `(left == right)`
%   left: `Some((['a', 'b'], Raw(35), Raw(36)))`,
%  right: `None`', src/learning.rs:2389:9
% stack backtrace:
%    0: backtrace::backtrace::libunwind::trace
%              at /cargo/registry/src/github.com-1ecc6299db9ec823/backtrace-0.3.46/src/backtrace/libunwind.rs:86
%    1: backtrace::backtrace::trace_unsynchronized
First, we see the example that Algorithm~\ref{alg:WFA} does not work properly if its $\textsc{Enclose-Column}$ is disabled.

\end{auxproof}

\begin{auxproof}
 \section{Column-Closedness for General Semirings}
 \label{sec:generalSemiring}
 We find the notion of column-closedness (used in the previous section) fundamental in automata learning. For example, it encompasses consistency.

 \begin{mytheorem}[general WFA learning]
 Let $\mathbb{S}$ be an arbitrary semiring.
 % Assume the following conditions.
 % \begin{enumerate}
 %  \item ...
 %  \item (Linear equations are solvable)
 % \end{enumerate}
 If a Hankel subblock is row- and column-closed, then it induces a faithful WFA.
 \end{mytheorem}

 (Show an algorithm here)

 \begin{mylemma}
 A mathematical result on finite generation of the Hankel matrix (Takamasa is not sure if this holds). 

 A back-up result: use PID structure, like in [van Heerdt], also for the column.
 \end{mylemma}

 As a consequence of the above mathematical results is the following ``correctness'' result.
 \begin{mytheorem}
 The above algorithm terminates if $\mathbb{S}$ is .... The resulting WFA is faithful to the Hankel subblock.
 \end{mytheorem}
\end{auxproof}

\section{Non-Termination and Non-Minimality}\label{subsec:mininalWFAs}
\begin{floatingfigure}[r]{0.17\textwidth}
\footnotesize
\!\!\!\!\!\!
\hspace{-3em}
\scalebox{.8}{    \begin{tabular}{c|cccc}
        $P\setminus S$ & $\epsilon$ & $b$ & $c$ & $\cdots$\\\hline
        $\epsilon$ &$0$ &$0$ &$0$ &$\cdots$\\
        $a$ & $0$& $1$& $2$&$\cdots$\\
        $aa$ & $0$& $2$& $4$& $\cdots$\\
        $\vdots$ & $\vdots$ & $\vdots$ & $\vdots$ & $\cdots$\\
        $a^n$ & $0$& $n$& $2\times n$& $\cdots$\\
        $\vdots$ & $\vdots$ & $\vdots$ & $\vdots$ & $\ddots$
    \end{tabular}
}
\\
\end{floatingfigure}
In general, Algorithm~\ref{alg:ours} does not terminate: even if the 
key conditions in \textsc{Enclose-Row} and \textsc{Enclose-Column}  are effectively checkable (Line~\ref{algline:rowcond} and the corresponding line in \textsc{Enclose-Column}), we may never get the Hankel subblock row- or column-closed.

To see it, consider the following example: for the WFA $B=(\{a,b,c\},\alpha_B,\beta_B,(B_\sigma)_{\sigma\in\Sigma})$ with
\begin{align}
    B_a = \begin{pmatrix}
        e_1\\
        1\otimes e_2\\
        2\otimes e_3
    \end{pmatrix},
    B_b = \begin{pmatrix}
        -\infty\\
        e_1\\
        -\infty
    \end{pmatrix},
    B_c = \begin{pmatrix}
        -\infty\\
        -\infty\\
        e_1
    \end{pmatrix},
\end{align}
where $-\infty$ means the row filled with $-\infty$ and $e_i$ means $-\infty$ with its $i$-th element replaced with $0$ for $i=1,2,3$, the Hankel matrix is as shown on the above right. It is not hard to see that no choice of finite rows generates the semimodule spanned by all the rows. Therefore Hankel subblocks are never row-closed from a certain moment on (specifically after $b,c\in S$), making Algorithm~\ref{alg:ours} diverge.

However, this does not mean that the semimodule spanned by all the rows is not finitely generated. In fact, since the Hankel matrix comes from the 3-state WFA $B$, there is a canonical choice of generators $r_1,r_2,r_3$ that span all the rows.
If we choose the row vectors $r_1,r_2,r_3\in \rmax^{\Sigma^{*}}$ by $r_i = (e_i B_{w_1}\dots B_{w_{\abs{w}}}\beta_B)_{w\in \Sigma^*}$ for $i=1,2,3$, then these row vectors span the submodule generated by all the rows of the Hankel matrix.

Once we find such row vectors, we can construct a WFA that is compatible with the Hankel matrix. This is by the following procedure. 
% (The problem is how to find such row vectors. \lstar-style algorithms search only in rows of the Hankel matrix, and for the max-plus semiring this is not enough, as shown by the above example $B$.)

For a Hankel matrix $H$ over the max-plus semiring, if there are vectors $r_1,\dots,r_n\in \rmax^{\Sigma^*}$ such that 
(1) the row $H(w, :)$ is expressed as $c_{w,1}\otimes r_1 \oplus \dots \oplus c_{w,n}\otimes r_n$ for any $w\in \Sigma^*$, and 
(2) there exists a homomorphism $\varphi_{\sigma}\colon \rmax^n\to \rmax^n$ for any $\sigma\in\Sigma$ such that $\varphi(c_{w,1},\dots,c_{w,n}) = (c_{w\sigma,1},\dots,c_{w\sigma,n})$ for any $w\in\Sigma^*$, then a WFA $A=(\Sigma, \alpha_A, \beta_A, (A_\sigma)_{\sigma \in \Sigma})$ which is consistent with $H$ can be constructed as follows:
\begin{itemize}
    \item $\alpha_A=(c_{\epsilon, 1},\dots,c_{\epsilon,n})$,
    \item $\beta_A = (r_1(\epsilon), \dots, r_n(\epsilon))^\top$,
    \item For $\sigma\in\Sigma$, $A_\sigma$ is the matrix induced from $\varphi_\sigma$.
\end{itemize}
The validity of the WFA is easy:
\begin{align}
    f_A(w) &= \alpha_A A_{w_1} \dots A_{w_m} \beta_A\\
    &= (c_{\epsilon, 1},\dots,c_{\epsilon,n})A_{w_1} \dots A_{w_m} \beta_A\\
    &= (c_{w_1,1},\dots,c_{w_1,n}) A_{w_2} \dots A_{w_m} \beta_A\\
    &= \dots\\
    &= (c_{w,1},\dots,c_{w,n})\beta_A\\
    &= (c_{w,1},\dots,c_{w,n}) \begin{pmatrix}
        r_1\\
        \vdots\\
        r_n
    \end{pmatrix} e_\epsilon^\top\\
    &= H(w,:)e_\epsilon^\top\\
    &= H(w,\epsilon).
% \\
%     &= f(w).
\end{align}

The problem is that \lstar-style algorithms (including Algorithm~\ref{alg:ours}) restrict its search for generators in \emph{rows of the Hankel matrix}. Lifting this restriction is at the cost of greater computational cost, and is a topic of future work.

This example is also an extreme witness that Algorithm~\ref{alg:ours} may not yield a minimal WFA: it yields infinitely many states while the minimal WFA has three states. 
% 類似の問題が工学のmax-plus systemの界隈で研究されていますが、計算量的に最小化は困難であると考えられています。
A similar problem of finding a minimal representation of max-plus linear systems has been studied  in control theory~\cite{612036,SCL:Schutter00}. In general, minimization of such systems is considered to be a computationally expensive problem. 
 % and it is considered to be computationally tough, so we guess solving this issue in the max-plus semiring is not easy.

Though it is not always possible to obtain the minimal WFA from Algorithm~\ref{alg:ours},
% We saw that it is sometimes impossible to obtain the minimal WFA from Algorithm~\ref{alg:ours}. However,
 we can still try to eliminate linearly dependent rows during the execution, so that we obtain a smaller WFA. In Appendix~\ref{appendix:bestEffortMinimization} we present a procedure and prove its correctness. The procedure consists of repeatedly discover redundant rows and removing them together with the relevant transitions.

\section{Conditional Termination of the Max-Plus WFA Learning}\label{sec:termination}
In the previous section, we proved that there is a case that Algorithm~\ref{alg:ours} does not terminate even for a weighted language generated from a WFA over the max-plus semiring.
% 先の章では、有限WFAから生成されたweighted languageにAlgorithm 1を適用しても、必ずしも停止しないことを示した。
In this section, we discuss a class of WFAs over the max-plus semiring for which Algorithm~\ref{alg:ours} terminates.
% この章では、停止することが保証できるようなmax-plus semiring上の有限WFAのクラスについて議論する。
WFAs whose elements are rational numbers and non-$(-\infty)$ form such a class.  Formally, the following theorem holds:
% エントリーに$-\infty$があらわれないようなWFAたちはそのようなクラスをなし、次の定理が成立する。
\begin{mytheorem}
    \label{thm:term}
    Let $A=(\Sigma, \alpha, \beta, (A_\sigma)_{\sigma\in \Sigma})$ be a WFA over the max-plus semiring. 
    If all the elements of the initial vector $\alpha$, the final vector $\beta$, and the transition matrices $A_\sigma (\sigma\in\Sigma)$ are rational numbers and non-$(-\infty)$, 
    and the principal solutions are used as solutions of max-plus linear systems in Algorithm~\ref{alg:ours},
    then Algorithm~\ref{alg:ours} applied to the language recognized by $A$ terminates.
    % max-plus semiring上のWFA Aについて、そのinitial vector, final vector,各tranistion matrixのどの要素にもεがあらわれず、かつその値たちが有理数であるとき、学習は必ず停止する。
\end{mytheorem}

% このクラスに属するWFAのことをrational WFAとよぶことにする+
We call such WFAs {\it rational}:
\begin{definition}[rational WFA]
    A WFA $A$ is called \emph{rational} if all the elements of the initial vector $\alpha$, the final vector $\beta$, and the transition matrices $A_\sigma (\sigma\in\Sigma)$ are rational and non-$(-\infty)$.
\end{definition}

% アルゴリズム中で使われる解についての条件は、定義ほげで述べたようにcomputationalに、またmathematicalに自然なものであり、またこの条件のおかげで停止性を証明することができる。
The condition on the solutions in Theorem~\ref{thm:term} is computationally and mathematically natural as we saw in Definition~\ref{def:principal_solution}, and it plays an important role in Proposition~\ref{prop510} to prove the termination.

% 証明のながれは次のようになる
The outline of the proof is as follows:
\begin{description}
    \item[Step 1] We see that the row contents of the Hankel matrix of a max-plus WFA are finite up to an equivalence relation, though they are infinite in general as we saw in Section~\ref{subsec:mininalWFAs}.   %はXで述べたように行の内容は無限通りであるが、ある同値関係のもとで有限であることを見る。(Secほげ)
    \item[Step 2] By the fact of Step 1, we prove that the addition of rows at Line~\ref{algline:add_p} cannot happen infinitely. %そのことを利用し、行が無限に追加されることがないことを証明し、Line Xが無限に実行されることはないことを示す。
    \item[Step 3] We prove that the addition of columns at Line~\ref{algling:add_s_by_eqq} cannot happen infinitely.  As this addition is caused only by equivalence queries and is not stopped by the checking of linear dependency, we cannot apply the technique of Step 2 straightforwardly, and we have to develop another technique. % equivalence query による反例の追加が無限に行なわれることはないことを示す。ここでも有限性を用いるが、反例の追加は行の線形独立性などとは関係なく行なわれるため、別のテクニックが必要になる。
    \item[Step 4] We prove that the addition of columns in {\sc Enclose-Column} (Line~\ref{algline:enclose_column}) cannot happen infinitely.  It is almost the same as Step 2, but slightly depends on Step 3. %2と同様に、今度は列が無限に追加されることがないことを示す。
    \item[Step 5] We wrap up Step 2-4, and prove Theorem~\ref{thm:term}. %の内容をまとめて、定理Xを証明する。
\end{description}

\subsection{Step 1: The finiteness of the row contents of the Hankel matrix of a max-plus WFA}
We prepare some notations and lemmas.
% この証明のために、いくらかterminologyとlemmaを準備する。normという語は~\cite{thebook}で用いられている。
\begin{mydefinition}[norm, scaled~\cite{thebook}]
    Let $v=(v_1,\dots,v_n)\in \rmax^n$ be a vector.
    The \emph{norm} $\norm{v}$ of $v$ is defined by $\max_i v_i$.
    A vector $v$ is \emph{scaled} if $\norm{v}=0$.
    For a vector $v\in \rmax^n$ such that at least one of the elements is not $-\infty$, $\scale(v)$ denotes $-\norm{v}\otimes v$.
\end{mydefinition}
For any vector $v$ satisfying the above condition, $\scale(v)$ is always scaled.
By applying $\scale$, we can focus on the differences of the elements of a vector.
We define the above notations for max-plus matrices similarly.
For example,
\begin{align*}
    \scale ((1, 2, 3)) = (-2, -1, 0),\quad \scale\begin{pmatrix}1 & 2\\3 & 4\end{pmatrix} = \begin{pmatrix}-3& -2\\-1& 0\end{pmatrix}.
\end{align*}

\begin{mydefinition}[height]
    For a max-plus vector $v=(v_1,\dots,v_n) \in \rmax^n$ such that none of $v_1,\dots,v_n$ is not $-\infty$, the \emph{height} $\height(v)$ denotes $(\max_i v_i) - (\min_i v_i)$.
    We define $\height$ similarly for max-plus matrices.
\end{mydefinition}
For example, $\height((1, 3, 2)) = 3 - 1 = 2$.
We can easily check that scalar multiplications in the max-plus semiring on vectors and matrices do not change their heights.

\begin{mylemma}
    \label{lem:xm}
    Let $M$ be a positive number (in $\real$), $d$ be a positive integer, and $X_M$ and $\mathbb{A}_M$ be
    \begin{align*}
        X_M &= \set{x\in (\rmax\setminus \set{-\infty})^d \mid \height(x)\le M},\\
        \mathbb{A}_M &= \set{A\in (\rmax\setminus \set{-\infty})^{d\times d} \mid \height(A)\le M}.
    \end{align*}
    For any matrix $A\in \mathbb{A}_M$ and any row vector $x\in X_M$, 
    $xA\in X_M$.
\end{mylemma}
\begin{myproof}
Since any scalar multiplication in the max-plus semiring does not change heights of vectors and matrices, without loss of generality,
we can assume that $x$ and $A$ are scaled, and $j$ be an index such that $x_j=0$.
All the elements of $x$ and $A$ are in $[-M, 0]$.
For any $i=1,\dots,d$, the $i$-th element of $xA$ is larger than or equal to $-M$:
\begin{align*}
    (xA)_i &= x_1\otimes A_{1i} \oplus \dots \oplus x_d\otimes A_{di}\\
    &\ge x_j\otimes A_{ji}\\
    &= 0\otimes A_{ji}\\
    &\ge -M.
\end{align*}
The non-positiveness of the elements of $xA$ is obvious.
Hence, all the elements of $xA$ are in $[-M, 0]$, and $xA \in X_M$ holds.\qed
\end{myproof}

\begin{mylemma}
    \label{lem:fin}
    Let $A=(\Sigma, \alpha, \beta, (A_\sigma)_{\sigma\in \Sigma})$ be a WFA over the max-plus semiring. 
    If $A$ is rational, the set $\set{\scale(\delta_A(w))\mid w\in \Sigma^*}$ is finite.
\end{mylemma}
\begin{myproof}
    Let $q$ be the GCD (in $\real$) of all the elements of $\alpha$, $\beta$, and $(A_\sigma)_{\sigma\in\Sigma}$, and 
    \begin{align*}
        M = \max\left(\height(\alpha), \height(\beta), \max_{\sigma\in\Sigma} \height(A_\sigma)\right).
    \end{align*}
    For any word $w\in \Sigma^*$, all the elements of $\delta_A(w)$ are multiples (in $\real$) of $q$.
    As $\delta_A(w) = \alpha\otimes A_{w_1}\otimes \dots \otimes A_{w_n}$, for any word $w$, $\height(\delta(w)) \le M$ by Lemma~\ref{lem:xm}.  This induces
    \begin{align*}
        \set{\scale(\delta(w)) \mid w\in \Sigma^*} \subset ([-M, 0] \cap q\mathbb{Z})^d.
    \end{align*}
    The left-hand side is finite, since the right-hand side is finite.
 \qed
\end{myproof}

These lemmas show a kind of finiteness of Hankel matrices generated from rational WFAs:
\begin{mylemma}
    \label{lem:fin_hankel}
    Let $A$ be a rational WFA.
    The contents of the rows of the Hankel matrix generated from $f_A$ are finite, i.e., the set of the rows
    \begin{align*}
        &\set{\text{($p$-th row of the Hankel matrix generated from $f_A$)}\in \rmax^{\Sigma^*} \mid p\in \Sigma^*}\\
        &= \set{(f_A(ps))_{s\in {\Sigma^*}} \in \rmax^{\Sigma^*} \mid p\in \Sigma^*}
    \end{align*}
    is finite up to the equivalence relation defined by
    \begin{align*}
        \text{($p$-th row)}\sim \text{($p'$-th row)} \iff
        \exists c\in \rmax \ \ \mathrm{s.t.}\ \ \text{($p$-th row)} = c\otimes \text{($p'$-th row)}.
    \end{align*}
    Dually, the contents of the columns of the Hankel matrix generated from $f_A$ are finite up to the equivalence relation defined by 
    \begin{align*}
        \text{($s$-th column)}\sim \text{($s'$-th column)} \iff
        \exists c\in \rmax \ \ \mathrm{s.t.}\ \ \text{($s$-th column)} = c\otimes \text{($s'$-th column)}.
    \end{align*}
\end{mylemma}
\begin{proof}
    The above set is 
    \begin{align*}
        \set{(f_A(s))_{s\in \Sigma}  \mid p\in \Sigma^*}
        &=
        \set{\left(\alpha \otimes \bigotimes_i A_{p_i} \otimes \bigotimes_i A_{s_i} \otimes \beta\right)_{s\in \Sigma}  \mid p\in \Sigma^*}\\
        &=
        \set{\alpha \otimes \bigotimes_i A_{p_i}  \mid p\in \Sigma^*} \otimes \left(\bigotimes_i A_{s_i} \otimes \beta\right)_{s\in \Sigma}\\
        &=
        \set{\delta_A(p) \mid p\in \Sigma^*}\otimes \left(\bigotimes_i A_{s_i} \otimes \beta\right)_{s\in \Sigma}.
    \end{align*}
    As the multiplicand is finite up to the equivalence relation $\sim$ by Lemma~\ref{lem:fin}, 
    the above set is also finite up to the equivalence relation.\qed
\end{proof}

\subsection{Step 2: The finiteness of calls of \textsc{Enclose-Row}}
\begin{myproposition}
    Let $A$ be a rational WFA.  The execution path of Algorithm~\ref{alg:ours} applied to $A$ satisfies the following properties.
    \begin{enumerate}
        \item The execution path does not visit Line~\ref{algline:add_p} infinitely many times.
        \item The execution does not call {\sc Enclose-Row} infinitely many times so that its return value ``$update?$'' is ``$updated$''.
    \end{enumerate}
    \label{prop56}
\end{myproposition}
\begin{myproof}
    We define $(P_i, S_i)$ as the $i$-th updated pair $(P, S)$ in the execution, where $i$ is a natural number.  Remark that even if just $P$ (or just $Q$) is updated, $(P_{i+1}, S_{i+1})$ is the next state of $(P_i, S_i)$ and $P_i = P_{i+1}$ (or $Q_i = Q_{i+1}$) can hold.

    We prove the first statement by contradiction.  Assume that the execution path is of infinite length and it visits Line~\ref{algline:add_p} infinitely many times.
    When Line~\ref{algline:add_p} is executed, $\mathtt{adding}$ is not $\mathtt{nil}$, as the line is in the else-branch of Line~\ref{algline:if_adding_nil}.
    This induces that Line~\ref{algline:set_adding} is also executed, as this line is the only line making $\mathtt{adding}$ non-$\mathtt{nil}$.
    Hence the condition on Line~\ref{algline:rowcond} is satisfied.  Hence, letting $i$ be the index of $(P_i, S_i)$ at a moment, there exists $p\in P_i$ and $\sigma \in \Sigma$ such that 
    \begin{align}
        xH(P_i, S_i) \neq H(\set{p\sigma}, S_i).  
    \end{align}
    at this step.  We call this {\it Fact~1} in this proof.
As the operation on $P$ is done only at Line~\ref{algline:add_p} and it adds just $p\sigma$, the set $P_i$ is prefix-closed for any $i$.
By Fact~1, for any $i$ and $p\in P_i$, $H(p, S_i)$ cannot be any linear combination of
\begin{align}
    \set{H(p', S_i) \mid \text{$p'$ is a proper prefix of $p$}}.
\end{align}
Hence $H(p, \Sigma^*)$ cannot be any linear combination of 
\begin{align}
    \set{H(p', \Sigma^*) \mid \text{$p'$ is a proper prefix of $p$}}.
\end{align}
We call this {\it Fact~2}.  As this addition occurs infinitely many times, defining $P_\infty = \bigcup_i P_i$, $P_\infty$ is an infinite set.

By Lemma~\ref{lem:fin_hankel}, $\set{H(p, \Sigma^*) \mid p\in \Sigma^*}$ is finite up to scaling.  Let $N$ be the index of the equivalence relation $\sim$ in Lemma~\ref{lem:fin_hankel}.  As $P_\infty$ is infinite and prefix-closed, there exists a word $p_N\in P_\infty$ of length $N$.
 Let $p_i$ be the prefix of $p_N$ of length $i$ for $i=0,\dots,N$.
 By the pigeonhole principle, there exists $i, j (0\le i < j \le n)$ such that $H(p_i, \Sigma^*) \sim H(p_j, \Sigma^*)$.
 Hence, there exists $c\in \rmax$ such that $H(p_j, \Sigma^*) = c\otimes H(p_i, \Sigma^*)$.
 This means $H(p_j, \Sigma^*)$ is a linear combination of 
 \begin{align}
    \set{H(p_k, \Sigma^*) \mid k=0,\dots,(j-1)},
 \end{align}
but this contradicts to Fact~2.  The first statement is proved.

    We prove the second statement.
Assume that {\sc Enclose-Row} is called infinitely many times so that its return value ``$update?$'' is ``$updated$''.
The return value ``$update?$'' being ``$updated$'' means Line~\ref{algline:set_updated} is visited in the execution, and it means Line~\ref{algline:add_p} is also visited.
Hence, Line~\ref{algline:add_p} is visited infinitely many times, but it contradicts to the first statement.
    \qed
\end{myproof}

\subsection{Step 3: The finiteness of call of equivalence queries}
\begin{mydefinition}[projection]
    Let $l$ and $r$ be natural numbers such that $l \le r$, and $x\in \rmax^n$ be a max-plus vector, where $n$ is a natural number including $\omega$ such that $r \le n+1$.
    The \emph{projection} of $x$ onto $(l, r)$ is defined by $\pi_{l, r}(x) = (x_l, x_{l+1}, \dots, x_{r-1})$.
\end{mydefinition}
For example, $\pi_{3, 6}(1, 2, 3, 4, 5, 6, 7) = (3, 4, 5)$.  Note that $\pi_{l, r}$ is max-plus linear.

\begin{mylemma}
    Let $N, M$ be natural numbers such that $N < M$, and $a_1, a_2, \dots, a_n, b\in \rmax^M$ be finite-valued max-plus vectors.
    If $x=(x_1,\dots,x_n)\in \rmax^n$ and $y=(y_1,\dots,y_n)\in \rmax^n$ satisfy two conditions: (1) the vector $x$ is a solution and a principal solution of 
    \begin{align}
        \pi_{1, N}(a_1) \otimes x_1 \oplus \dots \oplus \pi_{1, N}(a_n)\otimes x_n = \pi_{1, N}(b),
        \label{eq15}
    \end{align}
    (2) the vector $y$ is a solution and a principal solution of
    \begin{align}
        a_1 \otimes y_1 \oplus \dots \oplus a_n\otimes y_n = b,
        \label{eq16}
    \end{align}
    then, $x_i \ge y_i$ holds for any $i \in \set{1,\dots, n}$.

    Moreover, $x$ and $y$ satisfies a condition (3) $a_1\otimes x_1 \oplus \dots \oplus a_n\otimes x_n \neq b$ then,
    there exists $i\in \set{1,\dots,n}$ such that $x_i > y_i$.
    \label{lem58}
\end{mylemma}
\begin{proof}
    By applying $\pi_{1, N}$ to the both sides of Equation~\ref{eq16}, we get 
    \begin{align}
        \pi_{1, N}(a_1) \otimes y_1 \oplus \dots \oplus \pi_{1, N}(a_n)\otimes y_n = \pi_{1, N}(b).
    \end{align}
    This shows that $y$ is also a solution of Equation~\ref{eq15}.  As $x$ is the principal solution of Equation~\ref{eq15}, 
    $x_i \ge y_i$ holds for any $i\in \set{1,\dots,n}$.  The first statement is proved.
    From the first statement and the third condition, the second statement is proved.\qed
\end{proof}

The projection clips a part of a list of numbers, so $a_1$ in Lemma~\ref{lem58} can be regarded as an extension of $\pi_{1,N}(a_1)$.
Lemma~\ref{lem58} states how the solutions are updated from $x_1,\dots,x_n$ to $y_1,\dots,y_n$ when the vectors $\pi_{1,N}(a_1),\dots,\pi_{1,N}(a_n)$ are 
extended to $a_1,\dots,a_n$.

\begin{mylemma}
    Let $q$ be a positive rational number, $a_1,\dots,a_2,\dots, a_n, b \in \rmax^\omega$ be max-plus vectors of length $\omega$, and $N_1 < N_2 < \dots$ be a strictly increasing sequence of natural numbers.  If they satisfy the following three conditions: (1) all the elements of $a_1,\dots,a_n, b$ are multiples of $q$, (2) there exists a natural number $M$ such that $\max_i a_{ij} - \min_i a_{ij} \le M$ for any $j\in \set{1,\dots, N}$, and (3) an equation for $x_1^k, \dots, x_n^k \in \rmax$ 
    \begin{align}
        \pi_{1,N_k}(a_1)\otimes x_1^k \oplus \dots \oplus \pi_{1,N_k}(a_n)\otimes x_n^k = \pi_{1,N_k}(b)
        \label{eq19}
    \end{align}
    is solvable for any natural number $k$, then letting $x_1^k,\dots,x_n^k$ be the principal solutions of Equation~\ref{eq19}, there exists a natural number $k$ such that 
    \begin{align}
        \pi_{1, N_{k+1}}(a_1)\otimes x_1^k \oplus \dots \oplus \pi_{1,N_{k+1}}(a_n)\otimes x_n^k = \pi_{1,N_{k+1}}(b).
        \label{eq20}
    \end{align}
    \label{lem59}
\end{mylemma}
\begin{proof}
    We are proving this lemma by contradiction.  
    Assume that for any natural number $k$, Equation~\ref{eq20} does not hold (aiming at a contradiction).
We are proving that there exists $i\in \set{1, \dots, n}$ such that $x_i^1=x_i^2=\dots$.  We call this fact Fact~1 in this proof.
If it does not hold, for any $i$, there exists $j\in \nat$ such that $x_i^{j-1} \neq x_i^j$. 
By Lemma~\ref{lem58}, $x_i^{j-1} > x_i^j$ and $x_i^1\ge x_i^2\ge \dots$ hold, and $x_i^1 > x_i^j$ holds.
As the range of $i$ is finite, there exists $j\in \nat$ such that for any $i\in \set{1,\dots,n}$, $x_i^1 > x_i^j$ holds\footnote{From the discussion above, a natural number $j_i$ exists such that $x_i^{j_i-1} > x_i^{j_i}$.  Defining $j$ to be $\max_i j_i$ satisfies the condition. }.
A contradictory inequality below is induced\footnote{Lower indices are used so that $x_i$ means the $i$-th element of a vector $x$.}:
\begin{align*}
    b_1 &=(\pi_{1,{N_1}}(b))_1\\
    &= (\pi_{1,{N_1}}(a_1) \otimes x_1^1 \oplus \dots \oplus \pi_{1,{N_1}}(a_n) \otimes x_n^1)_1 \quad \text{(by the third assumption for case $k=1$)}\\
    &= (a_1 \otimes x_1^1 \oplus \dots \oplus a_n \otimes x_n^1)_1 \\
    &> (a_1 \otimes x_1^j \oplus \dots \oplus a_n \otimes x_n^j)_1 \quad \text{(by the property of $j$)}\\
    &= (\pi_{1,N_j}(a_1) \otimes x_1^j \oplus \dots \oplus \pi_{1,N_j}(a_n) \otimes x_n^j)_1 \\
    &=  (\pi_{1,N_j}(b))_1\quad \text{(by the third assumption for case $k=j$)}\\
    &=b_1 .
\end{align*}
Fact~1 is proved.  Without loss of generality, we can assume that $i=1$.  Let $x_1$ be $x_1^1=x_1^2=\dots$.

By the first assumption and the definition of the principal solution, $x_i^j$ is a multiple of $q$ for any $i$ and $j$.
By Lemma~\ref{lem58}, the assumption of proof by contradiction, for any $k\in \nat$, there exists $i\in \set{1,\dots,n}$ such that $x_i^k > x_i^{k+1}$.
If $n=1$, then $i=1$ must hold and the inequality contradicts to Fact~1, and the lemma is proved for the case $n=1$.
We assume $n\ge 2$ below.  As the range of $i$ is finite, there exists $i\in \set{2,\dots,n}$ such that a sequence $(x_i^1, x_i^2, \dots)$ decreases infinitely many times.
As the difference of one decrement is at least $q$, the sequence is unbounded below.  Without loss of generality, we can assume that $i=2$.  There exists $k\in \nat$ such that
\begin{align}
    x_2^k < x_1 - M.
    \label{eq27}
\end{align}
By the definition of principal solutions, there exists $j\in \set{1,\dots, N}$ such that
\begin{align}
    b_j = (x_2^k\otimes a_2)_j
    \label{eq28}
\end{align}
holds\footnote{By the definition of principal solutions, $x_2^k=\min_j (b_j-a_{2j})$.
By letting $j = \mathop{\mathrm{argmin}}_j (b_j-a_{2j})$, $x_2^k = b_j - a_{2j}$ holds.}.
A contradictory inequality below is induced:
\begin{align*}
    b_j &= (x_2^k \otimes a_2)_j \quad \text{(by Equation~\ref{eq28})}\\
    &= x_2^k\otimes a_{2j}\\
    &< (x_1-M)\otimes a_{2j}\quad \text{(by Equation~\ref{eq27})}\\
    &= x_1\otimes(-M)\otimes a_{2j}\\
    &\le x_1\otimes (-M)\otimes (a_{1j}+M) \quad \text{(by the second assumption)}\\
    &= x_1 \otimes a_{1j}\\
    &\le (a_1\otimes x_1^k \oplus \dots \oplus a_n \otimes x_n^k)_j \quad \text{(by the definition of $x_1$)}\\
    &= b_j \quad \text{(by definition of $x_1^k,\dots,x_n^k$)}.
\end{align*}
The lemma is proved for all the cases.\qed
\end{proof}
Remark that the indices $k$ and $k+1$ appear at the same time in Equation~\ref{eq20}; the principal solutions $x_1^k,\dots,x_n^k$ for Equation~\ref{eq19} are also solutions for Equation~\ref{eq19} whose indices are substituted from $k$ to $k+1$.   Lemma~\ref{lem59} states that there exists a moment such that the principal solutions $x_1^k,\dots,x_n^k$ can be reused for the solutions for the next extended equation when the vectors of the equations are extended infinitely many times.

\begin{myproposition}
    Let $A$ be a rational WFA.  If the principal solutions are used as solutions of max-plus linear systems in Algorithm~\ref{alg:ours}, the execution path of Algorithm~\ref{alg:ours} does not visit the else-clause at Line~\ref{algline:ce} infinitely many times.
    \label{prop510}
\end{myproposition}
\begin{proof}
    Assume that the execution path visits the else-clause infinitely many times.
    We define $(A_{\ext i}, P_i, S_i)$ as the $i$-th updated triple $(A_{\ext}, P, S)$ in the execution, where $i$ is a natural number.
    We use the same notation $(P_i, S_i)$ as Proposition~\ref{prop56}.
    As only Line~\ref{algline:add_p} manipulates $P$, there exists $I\in \nat$ such that $P_I = P_{I+1} = P_{I+2}=\dots$ by Proposition~\ref{prop56}.  Let $P_\infty = P_I$.

    At Line~\ref{algline:ce}, $f_{A_\ext}(w) \neq m(w)$ holds by the property of equivalence queries, and at Line~\ref{algline:extractWFA}, $f_{A_\ext}(w) = m(w)$ by the property of \textsc{Extract}.
    As the else-clause is visited infinitely many times, $A_\ext$ is updated infinitely many times.
    Let $I_1 < I_2 < \dots$ be a strictly increasing infinite sequence defined by sorting an infinite set
    \begin{align*}
        \set{i \in \nat \mid \text{$i \ge I$ and $A_{\ext i}\neq A_{\ext (i+1)}$}}.
    \end{align*}
    Remark that $A_{\ext I_1} \neq A_{\ext I_2} \neq \dots$ holds.
    As $P_{I_1}=P_{I_2}=\dots$ holds, the numbers of states, the initial vectors $\alpha$, and finial vectors $\beta$ of $A_{\ext I_1}, A_{\ext I_2}, \dots$ are the same by the construction discussed in Theorem~\ref{thm:main}.
    As $\Sigma$ and $P_\infty$ are finite, there exist $\sigma \in \Sigma$, $p\in P_\infty$, and an infinite subsequence $J_1, J_2,\dots$ of $I_1, I_2, \dots$ such that for any $j\in \nat$, the $p$-th row of the transition matrix of $A_{\ext J_j}$ for $\sigma$ is different from the $p$-th row of the transition matrix of $A_{\ext J_{j+1}}$ for $\sigma$, i.e., $(A_{\ext J_j})_\sigma (p, :) \neq (A_{\ext J_{j+1}})_\sigma (p, :)$.

    By the construction of Theorem~\ref{thm:main}, the $p$-th row of the transition matrix of $A_{\ext J_j}$ for $\sigma$ is the principal solution $x_j$ of the equation 
    \begin{align}
        x_j H(P_\infty, S_{J_j}) = H(\set{p\sigma}, S_{J_j}) 
    \end{align}
    for any $j\in \nat$.  As $(A_{\ext J_j})_\sigma (p, :) \neq (A_{\ext J_{j+1}})_\sigma (p, :)$ holds, $x_j \neq x_{j+1}$ also holds for any $j\in \nat$.
    By Lemma~\ref{lem58}, $x_j \ge x_{j+1}$ holds.
    In Algorithm~\ref{alg:ours}, as $S$ is extended so that $S$ keeps suffix-closed, $S_{J_1} \subseteq S_{J_2} \subseteq \dots$.
    By Lemma~\ref{lem:fin_hankel}, there exists $M\in \nat$ such that, for any $s\in \Sigma^*$, $\height(H(:, s)) \le M$.
    As $A$ is rational, there exists a positive rational number $q$ such that all the elements of the Hankel matrix are multiples of $q$.
    We use Lemma~\ref{lem59} with following settings:
    \begin{itemize}
        \item We assume the set of column indices $\Sigma^*$ are ordered by $\nat$ so that the first $\abs{S_{J_j}}$ elements are in $S_{J_j}$ for any $j\in \nat$.
        \item The vectors $a_1,\dots,a_n$ in the theorem are the rows of $H(P_\infty, \Sigma^*)$.
        \item The vector $b$ in the theorem is $H(\set{p\sigma}, \Sigma^*)$.
        \item The sequence $(N_i)_{i\in \nat}$ in the theorem is $(J_i)_{i\in \nat}$.
    \end{itemize}
    Then there exists $j\in \nat$ such that $x_j H(P_\infty, S_{J_{j+1}}) = H(\set{p\sigma}, S_{J_{j+1}})$.
    As $x_{j+1}$ is the principal solution of this equation regarding $x_j$ as its indeterminate variable, $x_j \le x_{j+1}$ holds in elementwise manner.  
    On the other hand, by the first statement of Lemma~\ref{lem58}, $x_j \ge x_{j+1}$ holds in elementwise manner.
    These inequalities shows that $x_j = x_{j+1}$, which is a contradiction.

    %% ギャップ: 日本語版コメント2だと埋まらないが、更新後のほうが解は小さくなっているはずなのに、それが最大解だっていうんだからそれはおかしいだろという気がする。
    \qed
\end{proof}

\subsection{Step 4: The finiteness of calls of \textsc{Enclose-Column}}
\begin{myproposition}
    Let $A$ be a rational WFA.  If the principal solutions are used as solutions of max-plus linear systems in Algorithm~\ref{alg:ours}, the execution path of Algorithm~\ref{alg:ours} applied to $A$ satisfies the following properties.
    \begin{enumerate}
        \item The execution path does not visit the line corresponding to Line~\ref{algline:add_p} in \textsc{Enclose-Column} infinitely many times.
        \item The execution does not call {\sc Enclose-Column} infinitely many times so that its return value ``$update?$'' is ``$updated$''.
    \end{enumerate}
    \label{prop511}
\end{myproposition}
\begin{proof}
    We use the notation $(P_i, S_i)$ in Proposition~\ref{prop56}.  
    By Proposition~\ref{prop56} and \ref{prop510}, there exists $i\in \nat$ such that $P_i = P_{i+1} = \dots$ and the update of $S_j$ to $S_{j+1}$ is caused only by the line corresponding to Line~\ref{algline:add_p} for any $j\ge i$.
    The rest of the proof is the dual of the Proposition~\ref{prop56}. 
    \qed
\end{proof}

\subsection{Step 5: Wrapping up}
\begin{proof}[of Theorem~\ref{thm:term}]
    By Proposition~\ref{prop56} and \ref{prop511}, \textsc{Enclose-Row} and \textsc{Enclose-Column} terminates, and they return $updated$ only finitely many times.
    Thus, the loop at Line~\ref{algline:extractStart} terminates, and \textsc{Extract} terminates.
    Assuming that the loop at Line~\ref{algline:mainloop} does not terminate, the else-clause at Line~\ref{algline:ce} is visited infinitely many times, but this cannot happen by Proposition~\ref{prop510}.
    Thus, the loop at Line~\ref{algline:mainloop} terminates.  All the loops in Algorithm~\ref{alg:ours} terminates, and Algorithm~\ref{alg:ours} terminates.
    \qed
\end{proof}

\section{Conclusions \& Future Work}
% \paragraph{Conclusions}
We proposed the column-closedness as the condition to construct a faithful WFA from a Hankel subblock, and proved the validity in a semiring-generic way without using the specific linear algebra over fields.
Based on this theory, we developed a new WFA learning method over general semirings including the max-plus semiring.
We proved that the column-closedness is necessary to ensure the faithfulness of constructed WFAs.
% The necessity to ensure the column-closedness in our method is discussed with a concrete example over the max-plus semiring.
We discussed the non-termination by a concrete example over the max-plus semiring, and identified the reason.  
% TODO: 私たちはどのようなときに私達のアルゴリズムが止まるか調べ、rationalなautomatonについて停止することを示しました。
% The potential applicability of the WFA learning over the max-plus semiring in the realistic setting is discussed by the noise-tolerant variant of our algorithm.
We proposed a noise-tolerant variant of our algorithm over the max-plus semiring towards the application of the WFA learning in a realistic setting.

% \paragraph{Future}

For future work in the theoretical direction, it would be interesting to find a class of max-plus automata with $-\infty$ that halts the learning.
Since rational automata have no $-\infty$ in the transition matrices, the weights of the states monotonically increase, and the proof in Section~\ref{sec:termination} took advantage of that monotonicity.
To use our current proof for other classes, it would be necessary to find classes that make the monotonicity hold or to extend monotonicity in some sense.
It would also be an interesting problem to investigate techniques to efficiently find (even in a heuristic way) a small set of generators of the Hankel matrix, as discussed in Section~\ref{subsec:mininalWFAs}. This would help to accelerate the learning and furthermore reduce the size of the resulting automaton.

In a practical direction, it would be useful to evaluate how often the equivalence query and membership query are issued in the various automaton learning algorithm variants.
In order to use automaton learning with max-plus for system identification, we would process the equivalence queries by comparing a candidate automaton made by the learning and the target system, and the membership queries by giving inputs to the target system.  As processing equivalence queries is more costly than processing membership queries, we are interested in the ratio of issued membership queries and equivalence queries.
In control theory, max-plus linear systems are used to model a system in reality.  As a max-plus linear system is almost the extension of a max-plus weighted automaton to accept continuous input values, there is a possibility that our method can be applied.

We have implemented van Heerdt et al.'s and our algorithms for the experiments to know practical performance, but the evaluation of the implementations is left open.
% We have implemented a program to measure it and conducted a preliminary experiment, but we found that the performances really depend on the problems and the general performance is still uncertain.
% Investigating the practical performances of van Heerdt et al.'s algorithm and our algorithm is also an important problem.  We have implemented a program to compare them, and found the performances really depend on the problems. The general tendency is still uncertain.  
The program is available on \url{https://github.com/ERATOMMSD/tropical_learning_public}.

\paragraph{Acknowledgments} 
This work is partially supported by
JST ERATO HASUO Metamathematics for Systems Design Project (No.\ JPMJER1603),
% The format of KAKENHI follows: https://www.jsps.go.jp/j-grantsinaid/16_rule/rule.html
JSPS KAKENHI Grant Numbers JP15KT0012, JP18J22498, JP19K20247, JP19K22842, and
JST-Mirai Program Grant Number JPMJMI18BA, Japan.

\bibliographystyle{splncs04}
\bibliography{aaai20}

\clearpage
\appendix
\section{Omitted Details}

% \subsection{Example Details for \S{}\ref{subsec:HeerdtEtAlDoesNotWork}}
% \label{appendix:exampleDeheerdt}
% The  WFA $A=(\Sigma, \alpha, \beta, (A_\sigma)_{\sigma\in \Sigma})$ we used for a membership oracle is as follows:
% $\Sigma=\set{a, b}$, $\alpha_A=(6, 11, 1)$, $\beta_A=(7, 0, 6)^\top$, 
% \begin{align*}
%     A_a = \begin{pmatrix}
%         2 & 3 & 1\\
%         2 & 0 & 9\\
%         3 & 0 & 8
%     \end{pmatrix},
%     A_b = \begin{pmatrix}
%         9 & 6 & 2\\
%         10 & 3 & 2\\
%         8 & 5 & 4
%     \end{pmatrix}.
% \end{align*}

% From the Hankel subblock $H_{(P,S)}$ shown in \S{}\ref{subsec:HeerdtEtAlDoesNotWork} (that is row-closed but not column-closed), we construct the following WFA:
% $\alpha_B = (0, -\infty, -\infty, -\infty), \beta_B=(13, 26, 35, 28)^\top$,
% \begin{align*}
%     \small
%     B_a = \begin{pmatrix}
%         8    &   0  &     -9   &   -2\\
%         16   &   8  &     -1   &   6\\
%         22   &   14 &     5    &   12\\
%         13   &   4  &     -5   &   2
%     \end{pmatrix},
%     B_b = \begin{pmatrix}
%         4   &    -4  &    -10 &    0\\
%         14  &   6    &   0    &   7\\
%         23  &    15  &    9   &    16\\
%         16  &    8   &    2   &    9
%     \end{pmatrix}.
% \end{align*}

\subsection{Best-Effort Minimization}\label{appendix:bestEffortMinimization}

When the rows of the Hankel subblock $H_{(P, S)}$ are weakly linearly dependent, we can reduce some rows and shrink the number of states of the constructed WFA in Theorem~\ref{thm:main}.
The reduction of the Hankel subblock is possible just by repeatedly sweeping the rows and find a row that can be expressed by the other rows and removing it.
It is known that the order of removing the row does not affect the number of the resulting weakly linear dependent rows (See Theorem 3.3.9 in~\cite{thebook}).

We assume that we get a WFA $A=(\Sigma, \alpha, \beta, (A_\sigma)_{\sigma\in \Sigma})$ as a result, and the Hankel subblock $H_{(P, S)}$ is weakly linearly dependent.
In this section, we index the matrices by numbers as the usual matrix theory, and $(-\infty,\dots,\banme{\text{$i$-th}}{0},\dots,-\infty)$ is abbreviated as $e_i$.
Without loss of generality, we can assume that the first $n$ number of rows of $H_{(P, S)}$ are weakly linearly independent.
By the definition of weak linear dependency, for $i=n+1, \dots, \abs{P}$, $H_{(P, S)}(i, :)$ can be expressed as $c_{i1}H_{(P, S)}(1, :) \oplus\dots\oplus c_{in}H_{(P, S)}(n, :)$ by coefficients $c_{i1},\dots,c_{in}\in \rmax$.  Hence, letting $C\in \rmax^{\abs{P}\times n}$ be 
\begin{align}
   C = 
    \kbordermatrix{
            & & & &\\
            &0 & -\infty & \cdots & -\infty\\
            &\vdots & \ddots & \ddots & \vdots\\
            \text{$n$-th \tiny $>$}& -\infty & \cdots & -\infty & 0\\
            \text{$(n+1)$-th \tiny $>$}& c_{n+1,1} & \cdots & \cdots & c_{n+1,n}\\
            & \vdots &\ddots &\ddots &\vdots \\
            \text{$\abs{P}$-th \tiny $>$}& c_{\abs{P},1} & \cdots & \cdots & c_{\abs{P},n}
        },
\end{align}
and $D\in \rmax^{n\times \abs{P}}$ be 
\begin{align}
    D = \begin{pmatrix}
        e_1\\
        \vdots\\
        e_n
    \end{pmatrix},
\end{align}
$H_{(P, S)} = CD H_{(P, S)}$ holds (remark that $CD$ is not the identity).  Then the WFA $B=(\Sigma, \alpha C, D\beta, (DA_\sigma C)_{\sigma \in \Sigma})$ yields the same language $f_A$, and its number of states is $n$, which is smaller than the number of states of $A$.  We give the sketch of the proof by showing $f_B(\sigma \tau) = f_A(\sigma \tau)$ for a word $\sigma \tau\in \Sigma^*$.   The idea is to reduce $CDH_{(P, S)}$ to $H_{(P, S)}$ by the relation above, and make the form of $CDH_{(P, S)}$ by using Lemma~\ref{lem:exchange}.
\begin{align}
    f_B(\sigma \tau) &=
    (\alpha C)(DA_\sigma C)(DA_\tau C)(D\beta )\\
    &=
    \alpha CDA_\sigma CDA_\tau CD H_{(P, S)}e_\epsilon^\top\\
    &=
    \alpha CDA_\sigma CDA_\tau H_{(P, S)}e_\epsilon^\top\\
    &=
    \alpha CDA_\sigma CD H_{(P, S)}B_\tau e_\epsilon^\top\\
    &=
    \alpha CDA_\sigma H_{(P, S)}B_\tau e_\epsilon^\top\\
    &=
    \alpha CD H_{(P, S)}B_\sigma B_\tau e_\epsilon^\top\\
    &=
    \alpha  H_{(P, S)}B_\sigma B_\tau e_\epsilon^\top\\
    &=
    \alpha A_\sigma A_\tau H_{(P, S)} e_\epsilon^\top\\
    &=
    \alpha A_\sigma A_\tau \beta\\
    &=
    f_A(\sigma \tau).
\end{align}

\end{document}